\documentclass[english]{IEEEtran}
\usepackage[T1]{fontenc}
\usepackage[latin9]{inputenc}
\usepackage{color}
\usepackage{babel}
\usepackage{url}
\usepackage{amsmath}
\usepackage{amsthm}
\usepackage{amssymb}
\usepackage{graphicx}
\usepackage[unicode=true,
 bookmarks=false,
 breaklinks=false,pdfborder={0 0 1},backref=false,colorlinks=false]
 {hyperref}
\hypersetup{
 colorlinks}

\makeatletter
\theoremstyle{plain}
\newtheorem{thm}{\protect\theoremname}
\theoremstyle{remark}
\newtheorem{rem}[thm]{\protect\remarkname}

\usepackage{algorithm}
\usepackage{algorithmic}
\usepackage{babel}

\providecommand{\remarkname}{Remark}
\providecommand{\theoremname}{Theorem}

\@ifundefined{showcaptionsetup}{}{%
 \PassOptionsToPackage{caption=false}{subfig}}
\usepackage{subfig}
\makeatother

\providecommand{\remarkname}{Remark}
\providecommand{\theoremname}{Theorem}

\begin{document}
\title{Channel Estimation and Reconstruction in Fluid Antenna System: Oversampling
is Essential}
\author{Wee Kiat New, \textit{Member, IEEE}, Kai-Kit Wong, \textit{Fellow,
IEEE}, Hao Xu, \textit{Member, IEEE,}\\
 Farshad Rostami Ghadi, \textit{Member, IEEE}, Ross Murch, \textit{Fellow,
IEEE}, and Chan-Byoung Chae, \textit{Fellow, IEEE} \thanks{The work of W. K. New, K. K. Wong, H. Xu, and F. R. Ghadi is supported
by the Engineering and Physical Sciences Research Council (EPSRC)
under grant EP/W026813/1. The work of R. Murch is supported by the
Hong Kong Research Grants Council Area of Excellence grant AoE/E-601/22-R.
The work of C.-B. Chae is supported by the Institute for Information
\& Communication Technology Planning \& Evaluation (IITP)/NRF grant
funded by the Ministry of Science and ICT (MSIT), Korea (No. 2021-0-02208,
No. 2021-0-00486).} \thanks{W. K. New (${\rm a.new@ucl.ac.uk}$), K.-K. Wong (${\rm kai\text{-}kit.wong@ucl.ac.uk}$),
H. Xu (${\rm hao.xu@ucl.ac.uk}$), and F. R. Ghadi (${\rm f.rostamighadi@ucl.ac.uk}$)
are with the Department of Electronic and Electrical Engineering,
University College London, London, WC1E 6BT, United Kingdom. K.-K.
Wong is also affiliated with Yonsei Frontier Lab., Yonsei University,
Seoul 03722, Korea.} \thanks{R. Murch (${\rm eermurch@ust.hk}$) is with the Department of Electronic
and Computer Engineering and Institute for Advanced Study (IAS), The
Hong Kong University of Science and Technology, Clear Water Bay, Hong
Kong SAR, China.} \thanks{C.-B. Chae (${\rm cbchae@yonsei.ac.kr}$) are with the School of Integrated
Technology, Yonsei University, Seoul, 03722, Korea.} \thanks{Corresponding author: Kai-Kit Wong.} \vspace{-5mm}
 }
\maketitle
\begin{abstract}
Fluid antenna system (FAS) has recently surfaced as a promising technology
for the upcoming sixth generation (6G) wireless networks. Unlike traditional
antenna system (TAS) with fixed antenna location, FAS introduces a
flexible component in which the radiating element can switch its position
within a predefined space. This capability allows FAS to achieve additional
diversity and multiplexing gains. Nevertheless, to fully reap the
benefits of FAS, obtaining channel state information (CSI) over the
predefined space is crucial. In this paper, we study the system with
a transmitter equipped with a traditional fixed antenna and a receiver
with a fluid antenna by considering an electromagnetic-compliant channel
model. We address the challenges of channel estimation and reconstruction
using Nyquist sampling and maximum likelihood estimation (MLE) methods.
Our analysis reveals a fundamental tradeoff between the accuracy of
the reconstructed channel and the number of estimated channels, indicating
that half-wavelength sampling is insufficient for perfect reconstruction
and that oversampling is essential to enhance accuracy. Despite its
advantages, oversampling can introduce practical challenges. Consequently,
we propose a suboptimal sampling distance that facilitates efficient
channel reconstruction. In addition, we employ the MLE method to bound
the channel estimation error by $\epsilon$, with a specific confidence
interval (CI). Our findings enable us to determine the minimum number
of estimated channels and the total number of pilot symbols required
for efficient channel reconstruction in a given space. Lastly, we
investigate the rate performance of FAS and TAS and demonstrate that
FAS with imperfect CSI can outperform TAS with perfect CSI. In contrast
to existing works, we also show that there is an optimal fluid antenna
size that maximizes the achievable rate when considering the energy
and bandwidth overheads for full CSI acquisition. 
\end{abstract}


\vspace{-5mm}

\section{Introduction}

\IEEEPARstart{F}{luid} antenna system (FAS) represents the new paradigm
of shape-and-position-flexible antenna technologies that benefit from
the new forms of reconfigurable antennas such as liquid-based antennas
\cite{dash2023selection,9899974}, reconfigurable radio-frequency
(RF) pixels \cite{1367557,9785489} or mechanical movable antenna
structures \cite{8060521}. The concept of FAS was originally brought
into wireless communication systems in 2020 \cite{Wong-ell2020} and
its application to the upcoming sixth-generation (6G) was discussed
in \cite{wong2022bruce,9770295,10480333}. Unlike a traditional antenna
system (TAS), where the antenna location is fixed, FAS refers to any
antenna structures that can intelligently adapt their shape and position
to reconfigure the gain, radiation patterns, and other radiation characteristics
to enhance communication performance. Recent articles in \cite{Shen-tap_submit2024,Zhang-pFAS2024}
have reported experimental results on FAS.

Recent efforts have treated FAS as a new degree of freedom (DoF) which
is dictated by its size and the number of flexible positions (a.k.a.~ports).
Early attempts have been to investigate the fundamental performance
limits of FAS, revealing considerable gains over TAS \cite{9264694,10103838,ramirez2024new,10130117,alvim2023performance,lopez2023novel}.
Most recently, the diversity and multiplexing gain for the multiple-input
multiple-output (MIMO) FAS setup was also understood \cite{10303274}.
Promising results by optimizing FAS in different setups can be found
in \cite{10354059,10354003,xu2023capacity,ISAC_FAS}. Furthermore,
there has been great interest in combining FAS with other technologies,
such as reconfigurable intelligent surfaces \cite{ghadi2024performance},
physical layer security \cite{10092780,Osorio-tvt2024}, non-orthogonal
multiple access \cite{10167904,10318134}, full-duplex radios \cite{10184308},
and index modulation \cite{zhu2023index,Xu-2024,Yang-2024}. A more
complete list of recent works is included in \cite{zhu2024historical}.

However, most existing studies rely on full channel state information
(CSI) to fully exploit the benefits of FAS. Recent advances in CSI
acquisition have greatly benefited from sophisticated signal processing
methods such as array processing \cite{8794743}, machine learning
\cite{8272484}, the Bayesian learning framework \cite{9931521},
compressive sensing \cite{5954192} and etc. Exploitation of channel
statistics, sparsity, and correlation has also enhanced channel estimation
performance by a lot \cite{9732214,9180053,10262375}. These techniques,
nonetheless, focus on scenarios with fixed radiating element locations,
thereby often limiting the number of channel estimations. In contrast,
in FAS scenarios, acquiring the CSI of a predefined space (i.e., full
CSI) is more challenging as it involves both channel estimation and
reconstruction (either interpolation or extrapolation, or both).

To obtain the full CSI, latest research has delved into the channel
estimation and reconstruction problem for FAS. For example, in \cite{10184308},
the authors devised a sequential linear minimum mean square error
(MMSE)-based channel estimation method for large-scale fluid antenna-aided
full-duplex cellular networks whereas \cite{10495003} employed deep
learning techniques to reconstruct the channels of the unobserved
ports based on the estimated channels from a few known ports. Besides,
\cite{10018377} and \cite{Eskandari-2024} used recurrent neural
networks (RNNs) and conditional generative adversarial networks (cGANs),
respectively, to tackle the channel reconstruction and port selection
issues in FAS. In scenarios where the channel is sparse, \cite{10375559}
illustrated that low-sample-size sparse channel reconstruction techniques
are effective for obtaining the full CSI. Additionally, a successive
Bayesian reconstructor was introduced in \cite{zhang2024successive,zhang2023successive}
with kernel-based sampling and regression techniques, and compressed
sensing methods were utilized in \cite{10236898}.

Collectively, these studies indicate that traditional channel estimation
schemes originally designed for TAS are inadequate for FAS, as they
would incur unbearable pilot overhead due to the possibility of positioning
the radiating element at near-infinitely many locations. To reduce
the pilot overhead, state-of-the-art methods typically only estimate
the channels at a few locations and reconstruct the FAS channel over
a finite space based on these initial estimates. This can be achieved
by leveraging the spatial correlation of the channels, their sparsity,
or any other implicit structural characteristics. Despite considerable
efforts, the minimum number of estimated channels and the minimum
distance between these estimated channels required to perfectly reconstruct
the FAS channel over a given space remain unknown. This challenge
persists partly because existing channel models provide limited insights
for more fundamental analysis, especially for the continuous FAS case
(i.e., when the number of ports is infinite for a given space). Note
that most literature on FAS has observed that increasing the number
of estimated channels or decreasing the minimum distance between the
estimated channels typically enhances the accuracy of channel reconstruction.

However, recent findings in holographic MIMO and electromagnetic information
theory present a contrasting perspective. Specifically, the functional
DoF represents the minimum number of samples required to reconstruct
a given electromagnetic field \cite{zhu2022electromagnetic}. This
metric is closely tied to the transmission capacity of an electromagnetic
system, as the maximum number of complex values that can be transmitted
within a single electromagnetic channel is equivalent to the minimum
number of required samples to reconstruct a given electromagnetic
field \cite{zhu2022electromagnetic,bjornson2024towards}. It was thus
argued in \cite{zhu2022electromagnetic,bjornson2024towards,9798854,di2023electromagnetic}
that half-wavelength sampling suffices to perfectly reconstruct an
arbitrary electromagnetic field with any specified precision in the
far-field because the electromagnetic field is band-limited. This
argument is supported by analyses in the space-wave number domain
or through the Petersen-Middleton's sampling theorem, both of which
are closely related to the Nyquist sampling theorem. In the near-field,
however, sampling at intervals smaller than half a wavelength is considered
beneficial \cite{di2023electromagnetic}.

Unlike existing studies, this paper answers two key questions: What
is the minimum number of estimated channels and the minimum distance
between these estimated channels needed to perfectly reconstruct the
FAS channel over a given space? Furthermore, using the answers to
these questions, can FAS still outperform TAS? To this end, we consider
a scenario in which a transmitter equipped with a traditional antenna
serves a receiver equipped with a fluid antenna over an electromagnetic-compliant
channel model. Specifically, the electromagnetic-compliant channel
helps us understand that the channel is band-limited when an infinite
plane is considered. However, considering the properties of the antenna,
such as its size and shape, reveals that the channel impinging on
the receive antenna experiences spectral leakage, causing it to no
longer be band-limited. Additionally, the channel is affected by receiver
noise, further complicating the channel estimation and reconstruction
processes. Leveraging these properties, we address the channel estimation
and reconstruction problem in FAS using tractable methods such as
Nyquist sampling and maximum likelihood estimation (MLE).

Through these model and methods, we show that perfect reconstruction
is impossible, revealing a fundamental tradeoff between the accuracy
of the reconstructed channel and the number of estimated channels.
This highlights the necessity of oversampling in FAS, even in far-field
propagation scenarios. Furthermore, we identify the minimum number
of estimated channels and the suboptimal distance between these estimated
channels necessary for efficient reconstruction of the FAS channel
over a finite space. Afterwards we study the rate of FAS with the
channel estimation and reconstruction problem, illustrating that FAS
with imperfect channels still outperforms TAS despite the necessity
to estimate the CSI over a given space. Additionally, we demonstrate
that there is an optimal fluid antenna size that maximizes the achievable
rate. Our key contributions are summarized as follows: 
\begin{itemize}
\item We have developed an electromagnetic-compliant channel model for FAS
for the first time, covering both one-dimensional (1D) and two-dimensional
(2D) fluid antenna surfaces. This model is applicable to reconfigurable
pixel-based fluid antennas in which the radiating elements are adjusted
in discrete manner, as well as to movable antennas or liquid-based
fluid antennas, where the number of ports approaches infinity, allowing
the radiating element to be repositioned continuously. More importantly,
we have considered the effects of antenna size and shape, as well
as noise, on channel estimation and reconstruction. 
\item Utilizing the Nyquist sampling theorem, we demonstrate a fundamental
tradeoff between the accuracy of the reconstructed channel and the
number of estimated channels. This finding underscores the necessity
of oversampling in FAS to ensure accurate channel reconstruction.
Furthermore, by employing the MLE method, we illustrate that the error
of the estimated channels can be bounded within a specific confidence
interval (CI). Through MLE, we also prove that the error of imperfect
channels can be modeled as Gaussian noise only at positions where
channel estimation is performed. 
\item To strike a balance between the accuracy of the reconstructed channel
and the number of estimated channels, we propose a suboptimal sampling
distance. This strategy, combined with MLE, allows us to determine
the minimum number of estimated channels needed for a given space
and obtain the number of pilot symbols required to efficiently estimate
and reconstruct the FAS channel. 
\item We present comprehensive simulation results that demonstrate the practicality
and efficiency of estimating and reconstructing the FAS channel on
both 1D and 2D fluid antenna surfaces. Additionally, our results highlight
the relationships between various parameters, such as sampling distance,
number of pilot symbols per sub-block, signal-to-noise ratio (SNR)
and CI. More importantly, we show that FAS with imperfect CSI can
outperform TAS with perfect CSI in a point-to-point scenario, and
there exists an optimal fluid antenna size that maximizes the achievable
rate after considering the overheads. 
\end{itemize}
Note that FAS can be conceptually viewed as a holographic MIMO system
with antenna selection. The reconstructed electromagnetic field can
be interpreted as the reconstructed FAS channel (i.e., channels over
a continuous space with infinitely many ports), while the sampling
points in the electromagnetic field can be seen as the estimated channels
in FAS. However, it is crucial to distinguish that an electromagnetic
field is proven to be band-limited (half-wavelength sampling sufficient
for perfect reconstruction) and the estimated FAS channels are affected
by noise. Leveraging ports that are less than half a wavelength apart
can also be interpreted as oversampling \cite{ramirez2024new}.

The remainder is organized as follows. Section~\ref{sec:emmodel}
presents the electromagnetic-compliant channel model for FAS. In Section~\ref{sec:sampling},
we discuss the channel estimation and reconstruction methods while
deriving important parameters, relationships, tradeoffs, and rates.
Simulation results are provided in Section~\ref{sec:results}. Finally,
the conclusion is drawn in Section~\ref{sec:conclude}.

\vspace{-1mm}

\section{Electromagnetic-Compliant Channel Model}

\label{sec:emmodel}

\begin{figure}
\includegraphics[width=8.5cm]{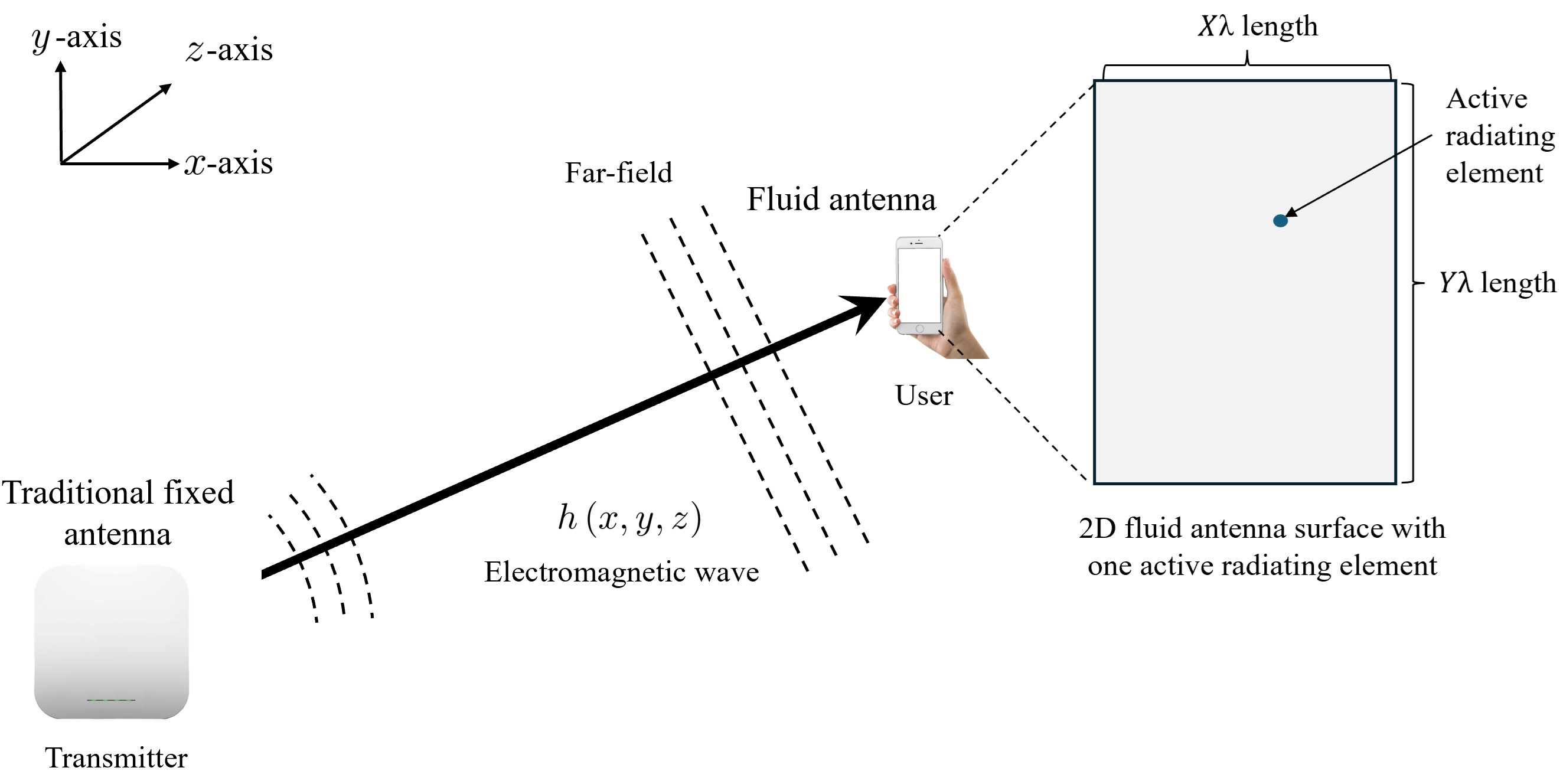}\caption{\label{fig:P2P_FAS}A point-to-point scenario with a traditional fixed
antenna at the transmitter and a 2D FAS at the receiver.}
\vspace{-5mm}
 
\end{figure}

As shown in Fig.~\ref{fig:P2P_FAS}, this paper considers a point-to-point
scenario in which a transmitter serves a receiver. We assume that
the transmitter is equipped with a traditional fixed antenna, while
the receiver employs a fluid antenna. Unless otherwise stated, we
assume that the fluid antenna surface is 2D and has a size of $X\lambda\times Y\lambda$
where $\lambda$ denotes the carrier wavelength. Furthermore, the
fluid antenna has only one active radiating element. In liquid-based
FAS \cite{9899974}, the radiating element can move to any location
on the 2D fluid antenna surface. In the case of a pixel-based antenna
\cite{9785489}, this means that certain pixels can be activated to
form an active radiating element. Compared to existing works, this
paper considers an electromagnetic-compliant channel model to gain
better understanding into the channel estimation and reconstruction
problem.\footnote{It is worth highlighting that \cite{8437634,9110848} are two pioneering
works that established electromagnetic-compliant channel models for
holographic MIMO. Motivated by these remarkable contributions, this
paper adopts a similar approach to develop an electromagnetic-compliant
channel model for FAS, applicable to both 1D and 2D surfaces.}

According to \cite{8437634}, the electromagnetic nature of small-scale
fading requires that each realization channel, $h\left(x,y,z\right)$,
satisfies the scalar Helmholtz equation, where $\left(x,y,z\right)$
represents the Cartesian coordinate in a three-dimensional (3D) space.
This stipulates that 
\begin{equation}
\left(\nabla^{2}+\kappa^{2}\right)h\left(x,y,z\right)=0,\label{eq:1}
\end{equation}
where (\ref{eq:1}) is a second-order linear partial differential
equation with constant coefficients, $\nabla^{2}$ is the Laplace
operator, and $\kappa=2\pi/\lambda$ is the wavenumber. Based on their
Fourier properties, it is evident from (\ref{eq:1}) that the following
condition must be met: 
\begin{equation}
k_{x}^{2}+k_{y}^{2}+k_{z}^{2}-\kappa^{2}=0.\label{eq:2}
\end{equation}
Given $k_{x}$ and $k_{y}$, we know that 
\begin{equation}
k_{z}=\pm k_{z}\left(k_{x},k_{y}\right)=\pm\sqrt{\kappa^{2}-k_{x}^{2}-k_{y}^{2}}.\label{eq:3}
\end{equation}
By superposition, the general solution to (\ref{eq:1}) is given by
\cite{hildebrand1962advanced} 
\begin{multline}
h\left(x,y,z\right)=\underset{\mathcal{D}\left(\kappa\right)}{\iint}H^{+}\left(k_{x},k_{y}\right)e^{j\left(k_{x}x+k_{y}y+k_{z}\left(k_{x},k_{y}\right)z\right)}\\
+H^{-}\left(k_{x},k_{y}\right)e^{j\left(k_{x}x+k_{y}y-k_{z}\left(k_{x},k_{y}\right)z\right)}dk_{x}dk_{y},
\end{multline}
where $H^{+}\left(k_{x},k_{y}\right),H^{-}\left(k_{x},k_{y}\right)$
are unknown constants and $\mathcal{D}\left(\kappa\right)=\left\{ \left(k_{x},k_{y}\right)\in\mathbb{{R}}\;|\;k_{x}^{2}+k_{y}^{2}\leq\kappa^{2}\right\} $.\footnote{By assuming that $k_{x}^{2}+k_{y}^{2}\leq\kappa^{2}$, we are focusing
on the far-field propagation. In the near-field, where $k_{x}^{2}+k_{y}^{2}>\kappa^{2}$,
it is evident that oversampling is useful \cite{di2023electromagnetic}.
However, as we will show later in this paper, oversampling remains
essential even in the far-field.}

The spatial autocorrelation function is given as 
\begin{multline}
c_{h}\left(x,y,z\right)\\
=\mathbb{{E}}\left\{ h\left(x+x',y+y',z+z'\right)h^{*}\left(x',y',z'\right)\right\} .\label{eq:5}
\end{multline}
Consequently, the power spectral density in the wavenumber domain
can be derived as 
\begin{multline}
S_{h}\left(k_{x},k_{y},k_{z}\right)\\
=\iiint c_{h}\left(x,y,z\right)e^{-j\left(k_{x}x+k_{y}y+k_{z}z\right)}dxdydz.\label{eq:6}
\end{multline}
Using the inverse Fourier transform, the power spectral density in
the wavenumber domain can be converted back into the spatial autocorrelation
function in the spatial domain by 
\begin{multline}
c_{h}\left(x,y,z\right)=\frac{1}{\left(2\pi\right)^{3}}\iiint S_{h}\left(k_{x},k_{y},k_{z}\right)\times\\
e^{j\left(k_{x}x+k_{y}y+k_{z}z\right)}dk_{x}dk_{y}dk_{z}.\label{eq:7}
\end{multline}
Due to (\ref{eq:1}) and the translation-invariant property, applying
the Helmholtz operators to both sides of (\ref{eq:5}) results in
\begin{equation}
\left(\nabla^{2}+\kappa^{2}\right)c_{h}\left(x,y,z\right)=0.\label{eq:8}
\end{equation}
By applying the Fourier transform to both sides of (\ref{eq:8}),
then we obtain 
\begin{equation}
\left(k_{x}^{2}+k_{y}^{2}+k_{z}^{2}-\kappa^{2}\right)S_{h}\left(k_{x},k_{y},k_{z}\right)=0.\label{eq:9}
\end{equation}
This implies that $S_{h}\left(k_{x},k_{y},k_{z}\right)=0$ except
at the wavenumber support, which is defined as 
\begin{equation}
S=\left\{ \left(k_{x},k_{y},k_{z}\right)\in\mathbb{{R}}^{3}\;|\;k_{x}^{2}+k_{y}^{2}+k_{z}^{2}=\kappa^{2}\right\} .
\end{equation}

Since $k_{x}^{2}+k_{y}^{2}+k_{z}^{2}=\kappa^{2}$ and $k_{z}$ is
determined by $k_{x}$ and $k_{y}$, as shown in (\ref{eq:2}) and
(\ref{eq:3}), respectively, we can simplify $S_{h}\left(k_{x},k_{y},k_{z}\right)$
as 
\begin{align}
S_{h}\left(k_{x},k_{y}\right)= & A^{2}\left(k_{x},k_{y}\right)\delta\left(k_{x}^{2}+k_{y}^{2}+k_{z}^{2}-\kappa^{2}\right),\label{eq:10}
\end{align}
where $A^{2}\left(k_{x},k_{y}\right)$ is the non-negative real function.
Using $(\ref{eq:3})$, we can define 
\begin{equation}
A^{2}\left(k_{x},k_{y}\right)=\begin{cases}
A_{+}^{2}\left(k_{x},k_{y}\right),k_{z}\left(k_{x},k_{y}\right)\geq0,\\
A_{-}^{2}\left(k_{x},k_{y}\right),k_{z}\left(k_{x},k_{y}\right)<0,
\end{cases}\left(k_{x},k_{y}\right)\in\mathcal{{D}}.\label{eq:11}
\end{equation}
According to \cite{balakrishnan2003all}, the Dirac delta function
can be split into the sum of two delta functions as 
\begin{multline}
\delta\left(k_{x}^{2}+k_{y}^{2}+k_{z}^{2}-\kappa^{2}\right)=\frac{\delta\left(k_{z}-k_{z}\left(k_{x},k_{y}\right)\right)}{2k_{z}\left(k_{x},k_{y}\right)}\\
+\frac{\delta\left(k_{y}+k_{z}\left(k_{x},k_{y}\right)\right)}{2k_{z}\left(k_{x},k_{y}\right)}.\label{eq:12}
\end{multline}
Substituting (\ref{eq:11}) and (\ref{eq:12}) into (\ref{eq:10}),
and then taking the inverse Fourier transform with respect to (w.r.t.)
$k_{z}$, we obtain 
\begin{align}
\frac{1}{2\pi}\int & S_{h}\left(k_{x},k_{y}\right)e^{j\left(k_{z}z\right)}dk_{z}\\
 & =\frac{A_{+}^{2}\left(k_{x},k_{y}\right)e^{+j\left(k_{z}\left(k_{x},k_{y}\right)z\right)}}{4\pi k_{z}\left(k_{x},k_{y}\right)}\nonumber \\
 & +\frac{A_{-}^{2}\left(k_{x},k_{y}\right)e^{-j\left(k_{z}\left(k_{x},k_{y}\right)z\right)}}{4\pi k_{z}\left(k_{x},k_{y}\right)}\label{eq:14}\\
 & =\iint c_{h}\left(x,y,z\right)e^{-j\left(k_{x}x+k_{y}y\right)}dxdy.\label{eq:15}
\end{align}
Note that in (\ref{eq:14}), we can define 
\begin{equation}
S_{h}^{\pm}\left(k_{x},k_{y}\right)=\frac{A_{\pm}^{2}\left(k_{x},k_{y}\right)}{4\pi k_{z}\left(k_{x},k_{y}\right)},~\left(k_{x},k_{y}\right)\in\mathcal{D}\left(\kappa\right),\label{eq:16}
\end{equation}
as the plane-wave spectrum.

According to \cite{8437634}, the random small-scale fading effect
can be modeled as 
\begin{align}
h\left(x,y,z\right)= & \frac{1}{2\pi}\iint H\left(k_{x},k_{y}\right)\nonumber \\
 & \times e^{j\left(k_{x}x+k_{y}y+k_{z}\left(k_{x},k_{y}\right)z\right)}dk_{x}dk_{y},\\
= & \frac{1}{2\pi}\iint\sum_{i=\pm}\sqrt{S_{h}^{i}\left(k_{x},k_{y}\right)}W_{i}\left(k_{x},k_{y}\right)\nonumber \\
 & \times e^{j\left(k_{x}x+k_{y}y\pm k_{z}\left(k_{x},k_{y}\right)z\right)}dk_{x}dk_{y}\label{eq:18}\\
= & \frac{1}{2\sqrt{\pi\kappa}}\underset{\mathcal{\mathcal{D}}\left(\kappa\right)}{\iint}\sum_{i=\pm}\frac{e^{j\left(k_{x}x+k_{y}y\right)}}{\left(\kappa^{2}-k_{x}^{2}-k_{y}^{2}\right)^{1/4}}W_{i}\left(k_{x},k_{y}\right)\nonumber \\
 & \times e^{\pm k_{z}\left(k_{x},k_{y}\right)z}dk_{x}dk_{y}\label{eq:19}
\end{align}
where $H\left(k_{x},k_{y}\right)$ is the spectrum of $h\left(x,y\right)$.
Furthermore, (\ref{eq:19}) is obtained by considering a 3D isotropic
channel and assuming that $h\left(x,y,z\right)$ has a unit power,
which yields $A_{\pm}\left(k_{x},k_{y}\right)=\frac{2\pi}{\sqrt{\kappa}}$.
Also, $W_{i}\left(k_{x},k_{y}\right)$ in (\ref{eq:18})
and (\ref{eq:19}) are the white Gaussian random process with unit
variance, used to model the stochastic nature of the propagation.\footnote{In (\ref{eq:19}), it is evident that the $z$-direction only introduces
a phase difference. Therefore, no additional functional DoF is actually
introduced, even if a 3D antenna surface is considered. In other words,
a 3D antenna surface does not fundamentally enhance the functional
DoF when the channel is electromagnetically compliant.} With this, we can express the unknown constants as 
\begin{equation}
H^{\pm}\left(k_{x},k_{y}\right)=\begin{cases}
\sqrt{{\scriptstyle S_{h}^{i}\left(k_{x},k_{y}\right)}}W_{i}{\scriptstyle \left(k_{x},k_{y}\right)}, & {\scriptstyle \left(k_{x},k_{y}\right)\in\mathcal{D}\left(\kappa\right),}\\
0, & {\rm {otherwise}}.
\end{cases}
\end{equation}
If we let $z=0$, then (\ref{eq:19}) can be rewritten as 
\begin{multline}
h\left(x,y,0\right)=\frac{1}{2\pi}\iint\sum_{i=\pm}\frac{\sqrt{\frac{\pi}{k}}}{\left(\kappa^{2}-k_{x}^{2}-k_{y}^{2}\right)^{1/4}}W_{i}\left(k_{x},k_{y}\right)\\
\times e^{j\left(k_{x}x+k_{y}y\right)}dk_{x}dk_{y}.\label{eq:21}
\end{multline}
This implies 
\begin{equation}
S_{h}^{\pm}\left(k_{x},k_{y}\right)=\frac{\frac{\pi}{k}}{\left(\kappa^{2}-k_{x}^{2}-k_{y}^{2}\right)^{1/2}},~\left(k_{x},k_{y}\right)\in\mathcal{D}\left(\kappa\right),\label{eq:22}
\end{equation}
and we can further simplify (\ref{eq:21}) as 
\begin{multline}
h\left(x,y\right)=\frac{1}{2\pi}\iint\sum_{i=\pm}\sqrt{S_{h}^{i}\left(k_{x},k_{y}\right)}W_{i}\left(k_{x},k_{y}\right)\\
\times e^{j\left(k_{x}x+k_{y}y\right)}dk_{x}dk_{y}.\label{eq:23}
\end{multline}
To approximate $h\left(x,y\right)$, we divide the
integration interval of (\ref{eq:23}) uniformly with frequency spacing
$\Delta_{k_{x}}=\frac{2\pi}{L_{x}}$ and $\Delta_{k_{y}}=\frac{2\pi}{L_{y}}$
so that 
\begin{equation}
h\left(x,y\right)\approx\sum_{\bar{y}=-\frac{\kappa L_{y}}{2\pi}}^{\frac{\kappa L_{y}}{2\pi}-1}\sum_{\bar{x}=-\frac{\kappa L_{x}}{2\pi}}^{\frac{\kappa L_{x}}{2\pi}-1}\sum_{i=\pm}H_{\bar{x},\bar{y}}^{i}e^{j2\pi\left(\frac{\bar{x}}{L_{x}}x+\frac{\bar{y}}{L_{y}}y\right)},
\end{equation}
where $\left(\bar{x},\bar{y}\right)\in\mathcal{E}$,
\begin{equation}
\mathcal{E}=\left\{ \left(\bar{x},\bar{y}\right)\in\mathbb{Z}^{2}\left|\left(\frac{\bar{x}\lambda}{L_{x}}\right)^{2}+\left(\frac{\bar{y}\lambda}{L_{y}}\right)^{2}\leq1\right.\right\} ,
\end{equation}
and
\begin{multline}
H_{\bar{x},\bar{y}}^{i}=\\
\int_{\frac{2\pi\bar{y}}{L_{y}}}^{\frac{2\pi\left(\bar{y}+1\right)}{L_{y}}}\int_{\frac{2\pi\bar{x}}{L_{x}}}^{\frac{2\pi\left(\bar{x}+1\right)}{L_{x}}}\sqrt{\frac{S_{h}^{i}\left(k_{x},k_{y}\right)}{\left(2\pi\right)^{2}}}W_{i}\left(k_{x},k_{y}\right)dk_{x}dk_{y}.
\end{multline}
Here, $L_{x}$ and $L_{y}$ are the number of discretized intervals
in $k_{x}$ and $k_{y}$ dimensions, while $H_{\bar{x},\bar{y}}^{i}$
computes the corresponding $\left(\bar{x},\bar{y}\right)$ interval
of $H^{i}\left(k_{x},k_{y}\right)$. 

Note that $H_{\bar{x},\bar{y}}^{i}$ is a Gaussian
random variable with zero mean and variance $\sigma_{i,\bar{x},\bar{y}}^{2}$.
The variance $\sigma_{i,\bar{x},\bar{y}}^{2}$, can be computed as
\begin{align}
\sigma_{i,\bar{x},\bar{y}}^{2} & =\int_{\frac{2\pi\bar{y}}{L_{y}}}^{\frac{2\pi\left(\bar{y}+1\right)}{L_{y}}}\int_{\frac{2\pi\bar{x}}{L_{x}}}^{\frac{2\pi\left(\bar{x}+1\right)}{L_{x}}}\frac{S_{h}^{i}\left(k_{x},k_{y}\right)}{\left(2\pi\right)^{2}}dk_{x}dk_{y},\\
 & =\frac{1}{4\pi}\int_{\frac{\bar{y}\lambda}{L_{y}}}^{\frac{\left(\bar{y}+1\right)\lambda}{L_{y}}}\int_{\frac{\bar{x}\lambda}{L_{x}}}^{\frac{\left(\bar{x}+1\right)\lambda}{L_{x}}}\frac{\boldsymbol{1}_{\mathcal{\mathcal{D}}\left(1\right)}\left(k_{x},k_{y}\right)}{\left(1-k_{x}{}^{2}-k_{y}{}^{2}\right)^{1/2}}dk_{x}dk_{y},
\end{align}

From the above derivations, it can be verified that $h\left(x,y\right)$
is band-limited because $H\left(k_{x},k_{y}\right)\in\mathcal{D}\left(\kappa\right)$.
According to the Nyquist sampling theorem \cite{1697831}, if $h\left(x,y\right)$
contains no frequencies higher than $\kappa$ in either $k_{x}$ or
$k_{y}$ dimensions, then $h\left(x,y\right)$ can be perfectly reconstructed
by uniformly sampling the channel with $D_{x},D_{y}\leq\frac{2\pi}{2\kappa}=\frac{\lambda}{2}$
distance. This is also known as the Nyquist sampling rate.

It is known that when a 3D isotropic channel is observed over an infinite
line, the electromagnetic field is 2D. We can use
similar steps to show that\footnote{Note that $A_{\pm}\left(k_{x}\right)=2\sqrt{\pi}$ when $h\left(x,y\right)$
has a unit power.} 
\begin{equation}
h\left(x\right)\approx\sum_{l=-\frac{\kappa L_{x}}{2\pi}}^{\frac{\kappa L_{x}}{2\pi}-1}\sum_{i=\pm}H_{l}^{i}e^{j2\pi\frac{l}{L_{x}}x},\label{eq:30}
\end{equation}
where $l\in\mathcal{E}_{l}$,
\begin{equation}
\mathcal{E}_{l}=\left\{ l\in\mathbb{Z}\left|\left(\frac{l\lambda}{L_{x}}\right)^{2}\leq1\right.\right\} ,
\end{equation}
\begin{equation}
H_{l}^{i}=\int_{\frac{2\pi l}{L_{x}}}^{\frac{2\pi\left(l+1\right)}{L_{x}}}\sqrt{\frac{S_{h}^{i}\left(k_{x}\right)}{2\pi}}W_{i}\left(k_{x}\right)dk_{x},
\end{equation}
and its variance, $\sigma_{i,l}^{2}$, can be computed as 
\begin{equation}
\sigma_{i,l}^{2}=\frac{1}{\pi}\left[\arcsin\left(\frac{l+1}{L_{x}}\right)-\arcsin\left(\frac{l}{L_{x}}\right)\right].
\end{equation}
Similarly, $h\left(x\right)$ can be perfectly reconstructed by uniformly
sampling the signal with $D_{x}=\frac{\lambda}{2}$ distance.

\section{Channel Estimation and Reconstruction using Nyquist Sampling and
MLE}

\label{sec:sampling} 
\begin{figure}[t]
\begin{centering}
\includegraphics[scale=0.4]{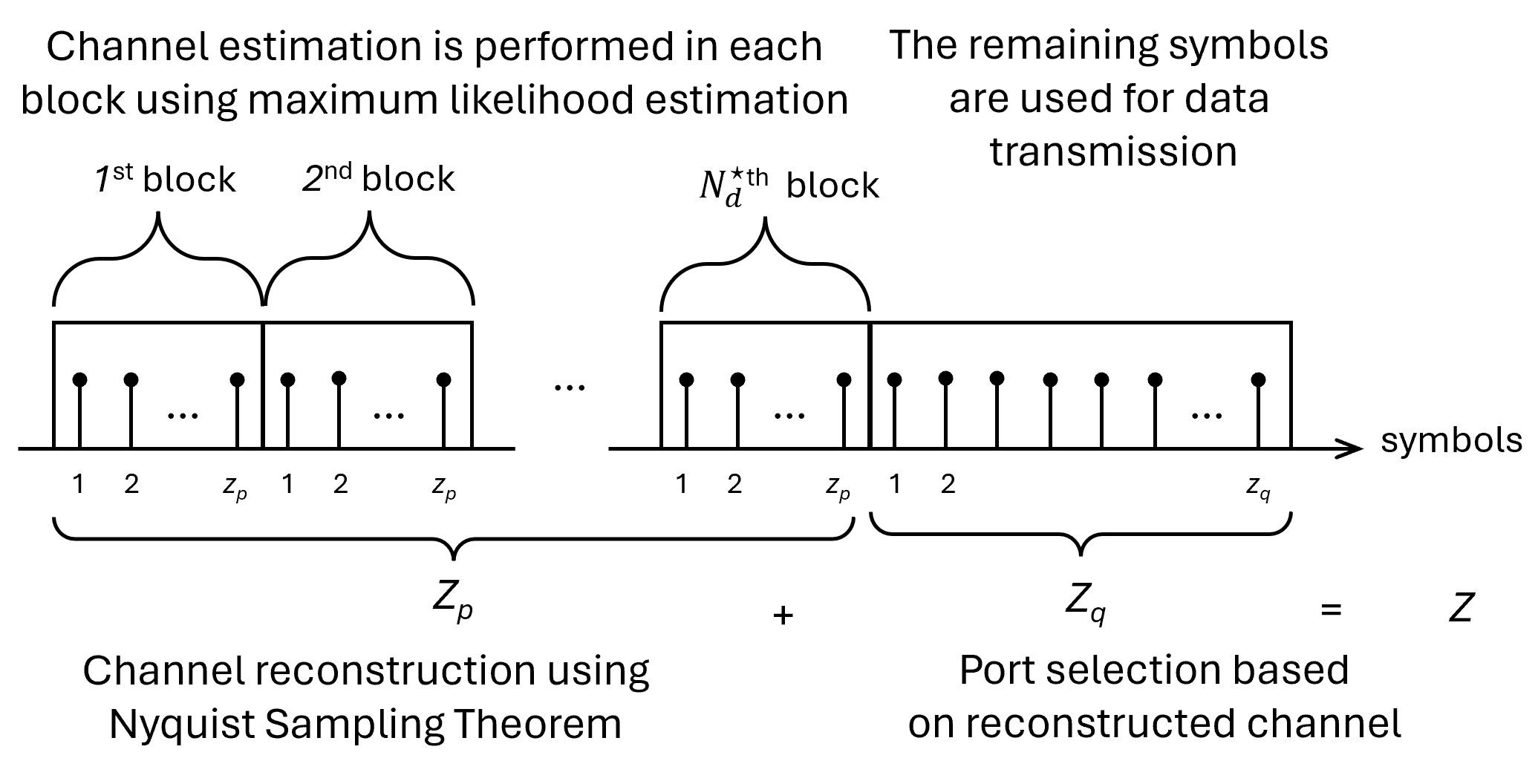} 
\par\end{centering}
\centering{}\caption{A schematic of the proposed channel estimation and reconstruction
processes within a coherence time.}
\label{fig:CER} \vspace{-3mm}
 
\end{figure}

As shown in Fig.~\ref{fig:CER}, we consider a slow flat fading channel
within a coherence time of $Z\gg0$ digital symbols such that $Z=Z_{p}+Z_{q}$,
where $Z_{p}$ and $Z_{q}$ are the total number of symbols used for
channel estimation and data transmission, respectively. The total
number of pilot symbols, $Z_{p}$, are further divided into $N_{d}^{\star}$
sub-blocks and each sub-block is dedicated to estimating the channel
at a predetermined location represented by $\left(n_{x}D_{x},n_{y}D_{y}\right)$,
where $n_{w}\in\mathbb{Z}$ and $D_{w}$ represents the sampling distance
in $w\in\left\{ x,y\right\} $ dimension. Each sub-block includes
$z_{p}$ pilot symbols for channel estimation purposes (e.g., using
MLE). For brevity, we focus our analysis on the real part of the channel
in this section as similar methodology can be applied to the imaginary
part in parallel.

Without loss of generality, let us treat $h_{{\rm FAS}}\left(x,y\right)$
as the real part of the channel of the FAS receiver at position $\left(x,y\right)$.
As it is infeasible to estimate $h_{{\rm FAS}}\left(x,y\right)$ for
$-\frac{X}{2}\lambda\leq x\leq\frac{X}{2}\lambda$ and $-\frac{Y}{2}\lambda\leq y\leq\frac{Y}{2}\lambda$
at every possible location, we leverage the Nyquist sampling method
for channel reconstruction. Using this method, we can then determine
the minimum value of $N_{d}^{\star}$ to efficiently reconstruct $\hat{h}_{{\rm FAS}}\left(x,y\right)$
from $\hat{h}_{{\rm FAS}}\left(n_{x}D_{x},n_{y}D_{y}\right),\forall n_{x,}n_{y}$.
Furthermore, we adopt the MLE method to estimate $\hat{h}_{{\rm FAS}}\left(n_{x}D_{x},n_{y}D_{y}\right)$.
This technique ensures that the estimation error for $\hat{h}_{{\rm FAS}}\left(n_{x}D_{x},n_{y}D_{y}\right)$
remains within a bound of $\varepsilon$, with a specific CI. This
approach ultimately enables us to determine the total number of pilot
symbols required for channel estimation and reconstruction in FAS.
Subsequently, we can analyze the effects of channel estimation and
reconstruction on the rate of FAS.

\subsection{2D Nyquist Sampling Method}

To reconstruct $\hat{h}_{{\rm FAS}}\left(x,y\right)$, we use the
2D Nyquist sampling method. As discussed in Section \ref{sec:emmodel},
$h\left(x,y\right)$ can be perfectly reconstructed by uniformly sampling
the channel with $\frac{\lambda}{2}$ distance since $H\left(k_{x},k_{y}\right)\in\mathcal{D}\left(\kappa\right)$.
Nevertheless, a FAS receiver can only observe $h\left(x,y\right)$
over a 2D fluid antenna surface, which has a finite space of $X\lambda\times Y\lambda$.
This is equivalent to multiplying $h\left(x,y\right)$ by a 2D rectangular
window function 
\begin{equation}
u\left(x,y\right)=\begin{cases}
1, & -\frac{X}{2}\lambda\leq x\leq\frac{X}{2}\lambda,-\frac{Y}{2}\lambda\leq y\leq\frac{Y}{2}\lambda,\\
0, & {\rm {otherwise}.}
\end{cases}
\end{equation}
As a result, the channel experienced at the FAS receiver is\footnote{Compared to \cite{8437634,9110848}, it is essential
to consider the antenna properties, such as the size and shape, in
FAS.} 
\begin{equation}
h_{{\rm FAS}}\left(x,y\right)=h\left(x,y\right)u\left(x,y\right).\label{eq:37}
\end{equation}
In the frequency domain, the spectrum of $h_{{\rm FAS}}\left(x,y\right)$
can be expressed as\footnote{Note that the terms ``frequency domain'' and ``wavenumber domain''
actually refer to the same conceptual domain. More concretely, the
term ``frequency domain'' is commonly used in signal processing,
whereas ``wavenumber domain'' is more frequently used in the context
of physics. In this paper, we use these terms interchangeably.} 
\begin{equation}
H_{{\rm FAS}}\left(k_{x},k_{y}\right)=H\left(k_{x},k_{y}\right)\circledast U\left(k_{x},k_{y}\right),\label{eq:38}
\end{equation}
where $\circledast$ is the 2D convolution operator. It can be easily
verified that 
\begin{equation}
U\left(k_{x},k_{y}\right)=\frac{\sin\left(\frac{k_{x}}{2}X\lambda\right)}{\frac{k_{x}}{2}}\frac{\sin\left(\frac{k_{y}}{2}Y\lambda\right)}{\frac{k_{y}}{2}},\label{eq:39}
\end{equation}
where the domain of $U\left(k_{x},k_{y}\right)$ spans over the entire
$\mathbb{R}^{2}$.

Now, substituting (\ref{eq:39}) into (\ref{eq:38}) suggests that
$H_{{\rm FAS}}\left(k_{x},k_{y}\right)$ is no longer band-limited.
As a matter of fact, $U\left(k_{x},k_{y}\right)$ leads to the spectral
leakage issue. This observation aligns with the uncertainty principle
\cite{landau1985overview}, which states that a spatially limited
signal cannot be simultaneously a band-limited signal and vice versa.
This means that perfect reconstruction of $h_{{\rm FAS}}\left(x,y\right)$
is impossible when the FAS receiver only observes $h\left(x,y\right)$
over a finite space of $X\lambda\times Y\lambda$. Thus, although
the Nyquist sampling rate of $\frac{\lambda}{2}$ is optimal for $h\left(x,y\right)$,
it is not optimal in the context of FAS, i.e., for $h_{{\rm FAS}}\left(x,y\right)$.
In other words, oversampling is essential in FAS to improve the accuracy
of the reconstructed channel.

This raises a fundamental tradeoff between the accuracy of reconstructed
channel and the number of estimated channels. Increasing the latter
can introduce hardware challenges and rate degradation. For instance,
a movable antenna or liquid-based fluid antenna would need to switch
positions more frequently, leading to higher energy consumption, while
a reconfigurable RF pixel-based fluid antenna might be subject to
stronger mutual coupling. In both cases, the computational complexity
of the CSI acquisition also increases. To strike a balance, we therefore
propose a suboptimal method for channel reconstruction that aims to
capture most of the energy of the mainlobe of $H_{{\rm FAS}}\left(k_{x},k_{y}\right)$.
In particular, the spectral leakage of the mainlobe of $U\left(k_{x},k_{y}\right)$
in the $k_{x}$- and $k_{y}$-dimensions are $\frac{2\pi}{X\lambda}$
and $\frac{2\pi}{Y\lambda}$, respectively \cite{lathi2005linear}.
Therefore, the spectral leakage of the mainlobe of $H_{{\rm FAS}}\left(k_{x},k_{y}\right)$
should extend to $\kappa+\frac{2\pi}{X\lambda}$ and $\kappa+\frac{2\pi}{Y\lambda}$,
respectively. This means that the suboptimal sampling distance can
be derived as 
\begin{align}
D_{x,0}^{\star} & =\frac{2\pi}{2\left(\kappa+\frac{2\pi}{X\lambda}\right)}=\frac{\lambda}{2+\frac{2}{X}},\label{eq:40}\\
D_{y,0}^{\star} & =\frac{2\pi}{2\left(\kappa+\frac{2\pi}{Y\lambda}\right)}=\frac{\lambda}{2+\frac{2}{Y}}.\label{eq:41}
\end{align}

Intuitively, we can improve the channel reconstruction by further
capturing the spectrum energy up to the $d$-th sidelobe at a cost
of smaller sampling distance.\footnote{In practice, the propagation may not be entirely isotropic. Sampling
at a smaller distance thus ensures a higher reconstruction accuracy.} Since the spectral spreading of the $d$-th sidelobe in the $k_{x}$-
and $k_{y}$-dimensions are simply $\left(d+1\right)\frac{2\pi}{X\lambda}$
and $\left(d+1\right)\frac{2\pi}{Y\lambda}$, respectively \cite{lathi2005linear},
the $d$-th suboptimal sampling distance can be expressed as 
\begin{align}
D_{x,d}^{\star} & =\frac{\lambda}{2+\frac{2}{X}\left(d+1\right)},\label{eq:42}\\
D_{y,d}^{\star} & =\frac{\lambda}{2+\frac{2}{Y}\left(d+1\right)}.\label{eq:43}
\end{align}
Note that (\ref{eq:40})--(\ref{eq:43}) can be regarded as the oversampling
rate since it is smaller than $\frac{\lambda}{2}$ \cite{ramirez2024new}.
Because the size of the 2D fluid antenna surface is fixed (e.g., $X\lambda\times Y\lambda$),
we can determine the number of estimated channels or sampling points
for a given space, denoted as $N_{d}^{\star}=N_{x,d}^{\star}N_{y,d}^{\star}$,
where 
\begin{align}
N_{x,d}^{\star} & =\left\lfloor \frac{X\lambda}{D_{x,d}^{\star}}\right\rfloor ,\label{eq:44}\\
N_{y,d}^{\star} & =\left\lfloor \frac{Y\lambda}{D_{y,d}^{\star}}\right\rfloor .\label{eq:45}
\end{align}
Very importantly, $N_{x,d}^{\star}$ and $N_{y,d}^{\star}$ are the
minimum numbers of estimated channels that include the spectral leakage
of the mainlobe or $d$-th sidelobe of the spectrum in the $x$ and
$y$ dimensions, respectively.

The sampled signal can be expressed as 
\begin{multline}
h_{{\rm FAS}}^{s}\left(x,y\right)=\sum_{n_{x}\in\mathbb{Z}}\sum_{n_{y}\in\mathbb{Z}}\hat{h}_{{\rm FAS}}\left(n_{x}D_{x,d}^{\star},n_{y}D_{y,d}^{\star}\right)\\
\times\delta\left(x-n_{x}D_{x,d}^{\star}\right)\delta\left(y-n_{y}D_{y,d}^{\star}\right),\label{eq:46}
\end{multline}
where $\hat{h}_{{\rm FAS}}(n_{x}D_{x,d}^{\star},n_{y}D_{y,d}^{\star})$
represents the estimated channel at location $(n_{x}D_{x,d}^{\star},n_{y}D_{y,d}^{\star})$.
In the frequency domain, we have 
\begin{align}
 & H_{{\rm FAS}}^{s}\left(k_{x},k_{y}\right)\nonumber \\
= & \sum_{n_{x}\in\mathbb{Z}}\sum_{n_{y}\in\mathbb{Z}}\hat{h}_{{\rm FAS}}\left(n_{x}D_{x,d}^{\star},n_{y}D_{y,d}^{\star}\right)e^{-j\left(k_{x}n_{x}D_{x,d}^{\star}+k_{y}n_{y}D_{y,d}^{\star}\right)}\nonumber \\
= & \frac{1}{D_{x,d}^{\star}D_{y,d}^{\star}}\sum_{l_{x}\in\mathbb{Z}}\sum_{l_{y}\in\mathbb{Z}}H_{{\rm FAS}}\left(k_{x}-\frac{2\pi}{D_{x,d}^{\star}}l_{x},k_{y}-\frac{2\pi}{D_{y,d}^{\star}}l_{y}\right).\label{eq:48}
\end{align}
Following the Nyquist sampling method, $\hat{h}_{{\rm FAS}}\left(x,y\right)$
can be reconstructed using a low-pass filter as 
\begin{equation}
\hat{H}_{{\rm FAS}}\left(k_{x},k_{y}\right)=H_{{\rm FAS}}^{s}\left(k_{x},k_{y}\right)f\left(k_{x},k_{y}\right),
\end{equation}
where 
\begin{equation}
f\left(k_{x},k_{y}\right)=\begin{cases}
1, & -\frac{2\pi}{D_{x,d}^{\star}}\leq k_{x}\leq\frac{2\pi}{D_{x,d}^{\star}},-\frac{2\pi}{D_{y,d}^{\star}}\leq k_{y}\leq\frac{2\pi}{D_{y,d}^{\star}},\\
0, & {\rm {otherwise}.}
\end{cases}
\end{equation}
Finally, we can obtain $\hat{h}_{{\rm FAS}}\left(x,y\right)$ for
$-\frac{X}{2}\lambda\leq x\leq\frac{X}{2}\lambda$ and $-\frac{Y}{2}\lambda\leq y\leq\frac{Y}{2}$
using the inverse Fourier transform as 
\begin{equation}
\hat{h}_{{\rm FAS}}\left(x,y\right)=\frac{1}{2\pi}\int\int\hat{H}_{{\rm FAS}}\left(k_{x},k_{y}\right)e^{-j\left(k_{x}x+k_{y}y\right)}dxdy.
\end{equation}
The normalized mean square error (NMSE) between $\hat{h}_{{\rm FAS}}\left(x,y\right)$
and $h_{{\rm FAS}}\left(x,y\right)$ can be computed as 
\begin{equation}
{\rm {NMSE}}=\frac{\int_{-\frac{Y}{2}\lambda}^{\frac{Y}{2}\lambda}\int_{-\frac{X}{2}\lambda}^{\frac{X}{2}\lambda}\left(\hat{h}_{{\rm FAS}}\left(x,y\right)-h_{{\rm FAS}}\left(x,y\right)\right)^{2}dxdy}{\int_{-\frac{Y}{2}\lambda}^{\frac{Y}{2}\lambda}\int_{-\frac{X}{2}\lambda}^{\frac{X}{2}\lambda}\left(h_{{\rm FAS}}\left(x,y\right)\right)^{2}dxdy}.
\end{equation}
The 1D counterpart can be derived in a similar fashion as will be
elaborated in Appendix I.

\subsection{MLE}

For brevity, let us denote $\left(\boldsymbol{D}\boldsymbol{n}\right)=\left(n_{x}D_{x},n_{y}D_{y}\right)$.
Suppose the radiating element of the FAS receiver is located at $\left(\boldsymbol{D}\boldsymbol{n}\right)$.
Then the received signal at symbol $z$ can be expressed as 
\begin{equation}
r_{z}\left(\boldsymbol{D}\boldsymbol{n}\right)=h_{{\rm FAS}}\left(\boldsymbol{D}\boldsymbol{n}\right)q_{z}+\omega_{z}\left(\boldsymbol{D}\boldsymbol{n}\right),~z=1,\ldots,z_{p},\label{eq:53}
\end{equation}
where $h_{{\rm FAS}}\left(\boldsymbol{D}\boldsymbol{n}\right)$ is
the unknown constant channel coefficient within the coherence time,
while $q_{z}$ and $\omega_{z}\left(\boldsymbol{D}\boldsymbol{n}\right)$
is, respectively, the pilot symbol and additive white Gaussian noise
(AWGN) during symbol $z$. For simplicity, we assume that $\left|q_{z}\right|^{2}=\frac{P}{2}$
and $\omega_{z}\left(\boldsymbol{D}\boldsymbol{n}\right)$ is an independent
Gaussian random variable with zero mean and variance of $\frac{N_{0}}{2}$,
where $P$ and $N_{0}$ is the transmit power and noise level, respectively.

To proceed, we can rewrite (\ref{eq:53}) as 
\begin{equation}
\boldsymbol{r}\left(\boldsymbol{D}\boldsymbol{n}\right)=\boldsymbol{q}h_{{\rm FAS}}\left(\boldsymbol{D}\boldsymbol{n}\right)+\boldsymbol{\omega}\left(\boldsymbol{D}\boldsymbol{n}\right).\label{eq:54}
\end{equation}
where 
\begin{equation}
\left\{ \begin{aligned}\boldsymbol{r}\left(\boldsymbol{D}\boldsymbol{n}\right) & =\left[r_{1}\left(\boldsymbol{D}\boldsymbol{n}\right)~\cdots~r_{z_{p}}\left(\boldsymbol{D}\boldsymbol{n}\right)\right]^{T},\\
\boldsymbol{q} & =\left[q_{1},\ldots,q_{z_{p}}\right]^{T}\\
\boldsymbol{\omega}\left(\boldsymbol{D}\boldsymbol{n}\right) & =\left[\omega_{1}\left(\boldsymbol{D}\boldsymbol{n}\right),\ldots,\omega_{z_{p}}\left(\boldsymbol{D}\boldsymbol{n}\right)\right]^{T}.
\end{aligned}
\right.
\end{equation}
In (\ref{eq:54}), $\boldsymbol{\omega}\left(\boldsymbol{D}\boldsymbol{n}\right)$
is a Gaussian random vector with zero vector mean and covariance of
$\frac{N_{0}}{2}\boldsymbol{I}$. Hence, the joint probability density
function (pdf) of $\boldsymbol{\omega}\left[n\right]$ is given by
\begin{align}
f_{\boldsymbol{\omega}}\left(w\right) & =\frac{1}{\pi N_{0}^{\frac{z_{p}}{2}}}\exp\left\{ -\sum_{z=1}^{z_{p}}\frac{\left(r_{z}\left(\boldsymbol{D}\boldsymbol{n}\right)-q_{z}h_{{\rm FAS}}\left(\boldsymbol{D}\boldsymbol{n}\right)\right)^{2}}{N_{0}}\right\} .\label{eq:55}
\end{align}
Note that (\ref{eq:55}) can be interpreted as the likelihood function
of the unknown parameter $h_{{\rm FAS}}\left(\boldsymbol{D}\boldsymbol{n}\right)$.
Thus, the log likelihood of $h_{{\rm FAS}}\left(\boldsymbol{D}\boldsymbol{n}\right)$
can be found as 
\begin{multline}
\mathcal{{L}}\left(\boldsymbol{r}\left(\boldsymbol{D}\boldsymbol{n}\right)|h_{{\rm FAS}}\left(\boldsymbol{D}\boldsymbol{n}\right)\right)\\
=-\frac{z_{p}}{2}\ln\left(\pi N_{0}\right)-\left[\sum_{z=1}^{z_{p}}\frac{\left(r_{z}\left(\boldsymbol{D}\boldsymbol{n}\right)-q_{z}h_{{\rm FAS}}\left(\boldsymbol{D}\boldsymbol{n}\right)\right)^{2}}{N_{0}^{z_{p}}}\right].\label{eq:56}
\end{multline}
The optimal condition for maximizing (\ref{eq:56}) w.r.t.~$h_{{\rm FAS}}\left(\boldsymbol{D}\boldsymbol{n}\right)$
(i.e., the MLE) is given by\footnote{Due to the electromagnetic properties, we assume that the distribution
of $h_{{\rm FAS}}\left(\boldsymbol{D}\boldsymbol{n}\right)$ is not
known as a priori. If the distribution of $h_{{\rm FAS}}\left(\boldsymbol{D}\boldsymbol{n}\right)$
is known, one can resort to the MMSE approach to further improve the
channel estimation performance.} 
\begin{align}
\hat{h}_{{\rm FAS}}\left(\boldsymbol{D}\boldsymbol{n}\right) & =\frac{\boldsymbol{q}^{H}\boldsymbol{r}\left(\boldsymbol{D}\boldsymbol{n}\right)}{\left\Vert \boldsymbol{q}\right\Vert ^{2}},\nonumber \\
 & =h_{{\rm FAS}}\left(\boldsymbol{D}\boldsymbol{n}\right)+\hat{\omega}_{z}\left(\boldsymbol{D}\boldsymbol{n}\right),
\end{align}
where $\hat{\omega}_{z}\left(\boldsymbol{D}\boldsymbol{n}\right)=\frac{\boldsymbol{q}^{H}}{\left\Vert \boldsymbol{q}\right\Vert ^{2}}\boldsymbol{\omega}\left(\boldsymbol{D}\boldsymbol{n}\right)$.
It can be easily verified that $\hat{h}_{{\rm FAS}}\left(\boldsymbol{D}\boldsymbol{n}\right)$
is also a Gaussian random variable. Specifically, it has a mean of
$h_{{\rm FAS}}\left(\boldsymbol{D}\boldsymbol{n}\right)$ and a variance
of $\left(z_{p}\;{\rm SNR}\right)^{-1}$. This implies that the estimator
is unbiased and the average mean square error is $\left(z_{p}\;{\rm SNR}\right)^{-1}$.
Using statistics, we can ensure that the estimation error of $\hat{h}_{{\rm FAS}}\left(\boldsymbol{D}\boldsymbol{n}\right)$
is less than $\varepsilon>0$ with a CI of 
\begin{equation}
\mathbb{{P}}\left\{ \left|\hat{h}_{{\rm FAS}}\left(\boldsymbol{D}\boldsymbol{n}\right)-h_{{\rm FAS}}\left(\boldsymbol{D}\boldsymbol{n}\right)\right|<\varepsilon\right\} ={\rm {erf}}\left(\varepsilon\sqrt{\frac{z_{p}\;{\rm SNR}}{2}}\right),\label{eq:62}
\end{equation}
where ${\rm {erf}\left(\cdot\right)}$ is the error function.\footnote{It is worth noting that outside the positions where channel estimation
is performed, the error depends on the reconstruction method and may
not necessarily follow a Gaussian distribution.} Since the error function is an increasing function, this suggests
that we can increase the CI either by increasing $\varepsilon$, $z_{p}$
or ${\rm SNR}$. For some fixed $\varepsilon$, ${\rm SNR}$, and
$\gamma$-CI, it is worth noting that we can find the minimum value
via a 1D searching method as 
\begin{equation}
z_{p}^{*}=\left\{ \min z_{p}\left|{\rm {erf}}\left(\varepsilon\sqrt{\frac{z_{p}\;{\rm SNR}}{2}}\right.\right)\geq\gamma\right\} .
\end{equation}
Thus, the total number of pilot symbols required for channel estimation
and reconstruction in FAS can be obtained as 
\begin{equation}
Z_{p}=z_{p}N_{d}^{\star}.
\end{equation}
Note that if $Z$ is not significantly larger than $Z_{p}$ (e.g.,
fast-fading channel), adjustments can be made to parameters such as
$\varepsilon$, ${\rm SNR}$, and $\gamma$-CI to reduce the required
symbols. Other options include employing other state-of-the-art channel
estimation methods or fixing the antenna location (i.e., reducing
FAS to TAS for a robust communication performance).\footnote{If the channel remains band-limited, the proposed method can be extended
to a MIMO setup with multiple fixed-position antennas at the transmitter.
The channel estimation and reconstruction for MIMO-FAS, however, remains
a complicated problem and requires more detailed analysis.}
\begin{rem}
In practice, the fast Fourier transform (FFT) and inverse fast Fourier
transform (IFFT) are typically employed. With a frequency resolution
of $F>N_{d}^{\star}$, the FFT and IFFT have a time complexity of
$\mathcal{O}\left(F\log_{2}\left(F\right)\right)$. In addition, the
low-pass filter has a time complexity of $\mathcal{O}\left(F\right)$,
and MLE for each location has a time-complexity
of $\mathcal{O}\left(z_{p}\right)$. Since $F$ and $z_{p}$ are independent
parameters, the overall time complexity of the proposed Nyquist sampling
and MLE method is $\mathcal{O}\left(F\log_{2}\left(F\right)+N_{d}^{\star}z_{p}\right)$. 
\end{rem}

\subsection{Data Transmission}

To maximize the data rate, we assume that the radiating element of
FAS is activated at the location with the maximum amplitude. In the
case of perfect CSI, the optimal position is 
\begin{equation}
\left(x^{*},y^{*}\right)=\arg\max_{\left(x,y\right)}\left\{ \left|h_{{\rm FAS}}\left(x,y\right)\right|^{2}\right\} .
\end{equation}
The received signal is expressed as 
\begin{equation}
r_{{\rm {perfect}}}=h_{{\rm FAS}}\left(x^{*},y^{*}\right)q+\omega,
\end{equation}
where $q$ is the random information signal and $\omega$ is the AWGN.
The capacity of the FAS receiver is given as 
\begin{equation}
R_{{\rm FAS}}^{*}=B\log_{2}\left(1+\left|h_{{\rm FAS}}\left(x^{*},y^{*}\right)\right|^{2}\frac{{\rm SNR}}{B}\right),
\end{equation}
where $B$ is the system bandwidth and ${\rm SNR}=\frac{P}{N_{0}}$.
In the case of imperfect CSI, it is necessary to estimate the channel
using the proposed Nyquist sampling and MLE methods. After that, the
FAS receiver can select its port based on the estimated and reconstructed
channel. The suboptimal position is 
\begin{equation}
\left(x^{\star},y^{\star}\right)=\arg\max_{\left(x,y\right)}\left\{ \left|\hat{h}_{{\rm FAS}}\left(x^{\star},y^{\star}\right)\right|^{2}\right\} .
\end{equation}

With imperfect CSI, we have the received signal \cite{841172} 
\begin{equation}
r_{{\rm {imperfect}}}=\hat{h}_{{\rm FAS}}\left(x^{\star},y^{\star}\right)q+\omega,\label{eq:65}
\end{equation}
and the achievable rate can be obtained as \cite{720551} 
\begin{equation}
R_{{\rm FAS}}^{\star}=B\left(1-\frac{N_{d}^{\star}z_{p}}{Z}\right)\log_{2}\left(1+\frac{\left|h_{{\rm FAS}}\left(x^{\star},y^{\star}\right)\right|^{2}{\rm SNR}}{\left|\epsilon\right|^{2}{\rm SNR}+B}\right),\label{eq:66}
\end{equation}
where the additional pre-log factor accounts for the remaining symbols
that can be used for data transmission per coherence time, and $\epsilon=h_{{\rm FAS}}\left(x^{\star},y^{\star}\right)-\hat{h}_{{\rm FAS}}\left(x^{\star},y^{\star}\right)$
accounts for the estimation error. Clearly, (\ref{eq:66}) shows that
there is a fundamental tradeoff between high data rate and accurate
channel estimation and reconstruction.

In addition, we can further consider FAS where the position of the
radiating element is fixed. This is equivalent to TAS, and the capacity
of TAS with perfect CSI is \cite{841172} 
\begin{equation}
R_{{\rm TAS}}=B\log_{2}\left(1+\left|h_{{\rm FAS}}\left(x,y\right)\right|^{2}\frac{{\rm SNR}}{B}\right),
\end{equation}
where $\left(x,y\right)$ is an arbitrary fixed position throughout
the time.

\begin{figure}
\begin{centering}
\subfloat[]{\begin{centering}
\includegraphics[scale=0.62]{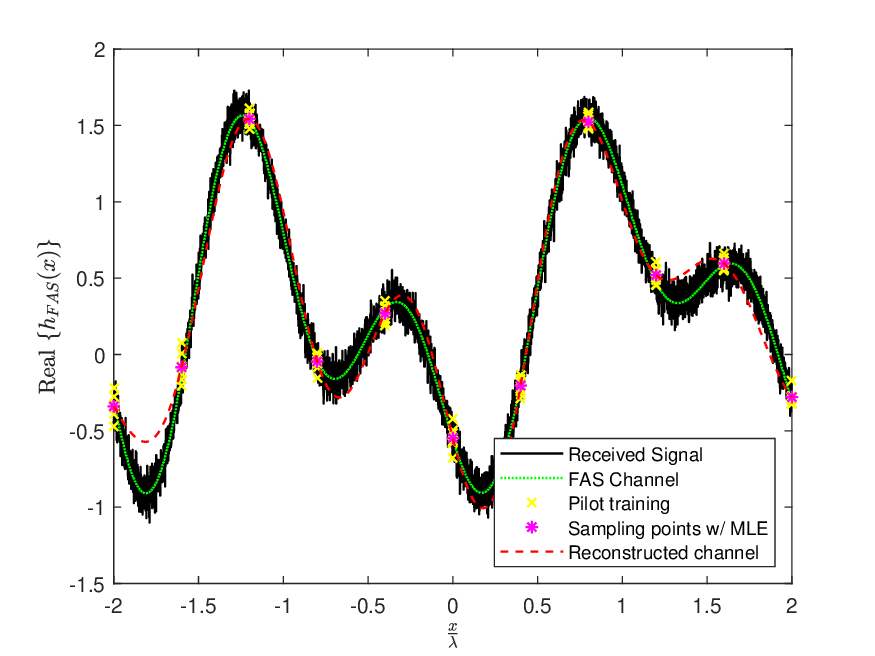} 
\par\end{centering}
}
\par\end{centering}
\begin{centering}
\subfloat[]{\begin{centering}
\includegraphics[scale=0.62]{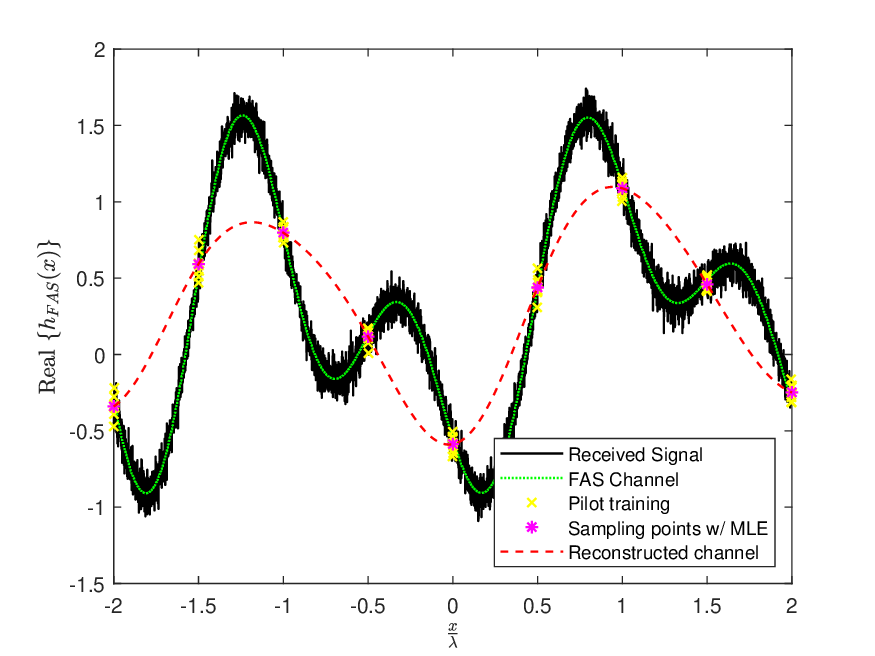} 
\par\end{centering}
}
\par\end{centering}
\centering{}\caption{The fundamental effects of different parameters on a 1D channel estimation
and reconstruction using: a) $D_{0}^{\star}$ sampling distance; and
b) $D=\frac{\lambda}{2}$ sampling distance.}
\label{fig:recon} \vspace{-4mm}
 
\end{figure}

\section{Results and Discussion}

\label{sec:results} In this section, we evaluate the performance
of the proposed Nyquist sampling and MLE methods for both channel
estimation and reconstruction, as well as rate. For simplicity, we
consider a symmetric setting; therefore, we denote $W$ as the size
and $D$ as the sampling distance per dimension. Unless otherwise
stated, we assume that $Z=1200$, $B=30~{\rm KHz}$, $z_{p}=7$, ${\rm {SNR}=20~dB}$,
$L_{x}=L_{y}=32$, $W=X=Y=2$, and $D=D_{0}^{\star}=D_{x,0}^{\star}=D_{y,0}^{\star}$.

\begin{figure}
\begin{centering}
\subfloat[]{\begin{centering}
\includegraphics[scale=0.62]{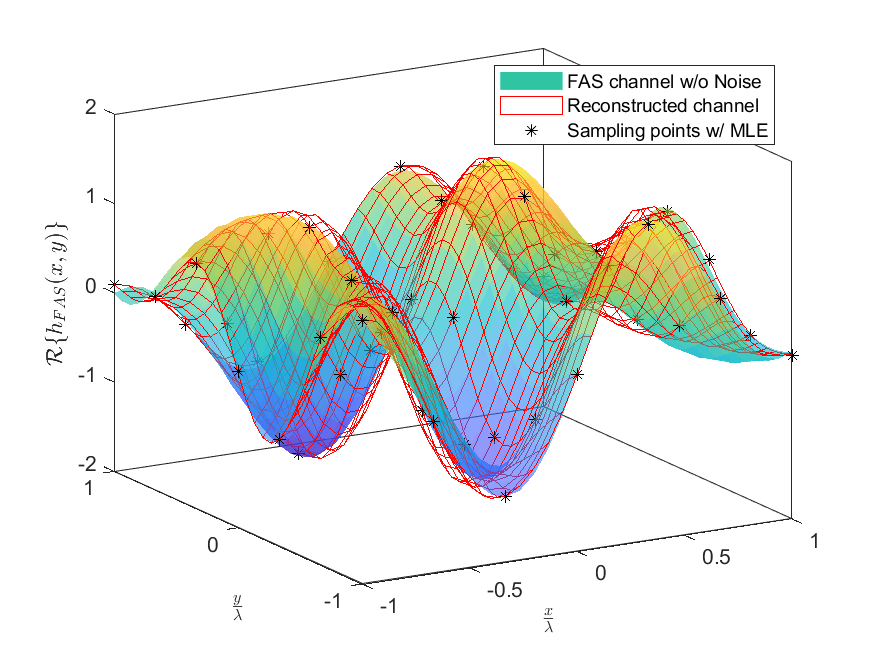} 
\par\end{centering}
}
\par\end{centering}
\begin{centering}
\subfloat[]{\begin{centering}
\includegraphics[scale=0.62]{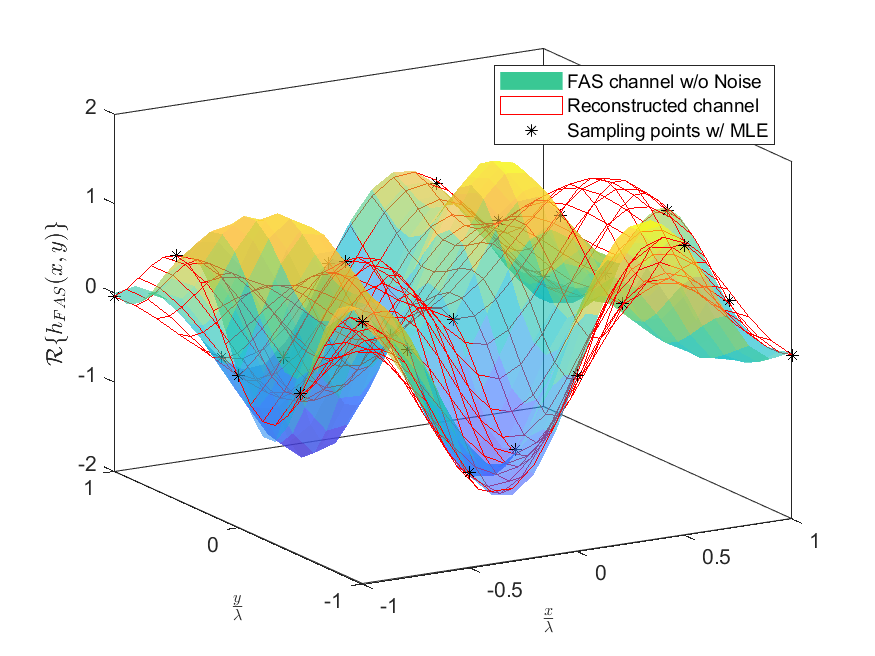} 
\par\end{centering}
}
\par\end{centering}
\centering{}\caption{The 2D channel estimation and reconstruction using: a) $D_{0}^{\star}$
sampling distance; b) $D=\frac{\lambda}{2}$ sampling distance.}
\label{fig:2D_Recon_spatial} \vspace{-4mm}
 
\end{figure}

We first explore the fundamental effects of different parameters on
the 1D channel estimation and reconstruction case, with $X=4$. For
the sake of brevity, we focus only on the real part, as similar behaviors
are observed in the imaginary part. As we can see in Fig.~\ref{fig:recon},
the SNR influences the variation of the received signals, while $z_{p}$
determines the amount of pilot training. When the SNR is low, a higher
$z_{p}$ can improve the accuracy of estimated channels or sampling
points by averaging out the noise effects with MLE. However, the accuracy
of the reconstructed channel strongly depends on both the accuracy
of MLE at sampling points and the sampling distance. As observed in
Fig.~\ref{fig:recon}(b), if the sampling distance is too large,
the inaccuracies in the reconstructed channel could be much more significant
than those caused by the SNR, (i.e., the variation of the received
signals). This suggests that the sampling distance is equally, if
not more, important than the accuracy of MLE at the sampling points.
Moreover, this result clearly highlights that a sampling distance
of $D=\frac{\lambda}{2}$ is insufficient to perfectly reconstruct
the channel.

\begin{figure}
\begin{centering}
\subfloat[]{\begin{centering}
\includegraphics[scale=0.62]{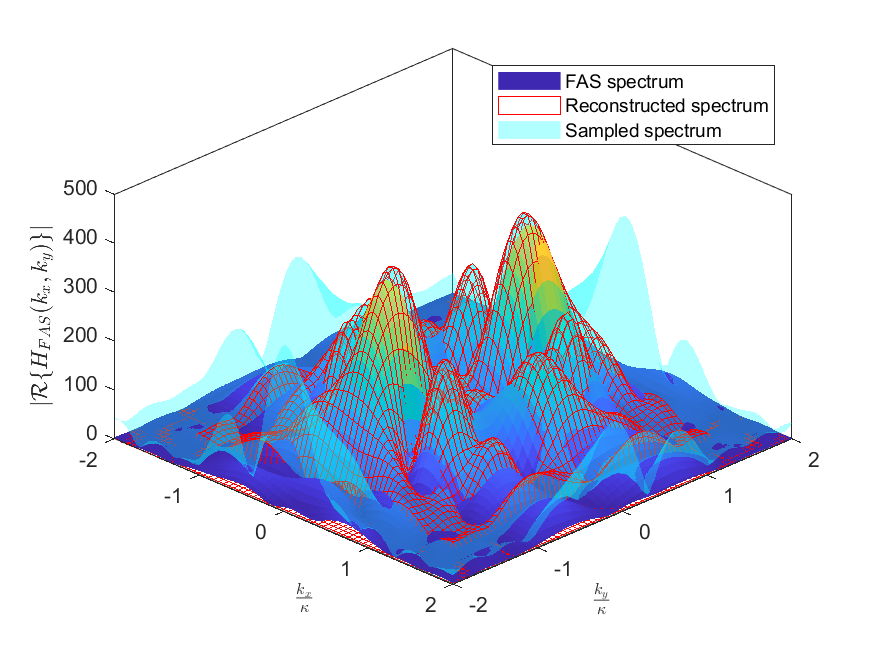} 
\par\end{centering}
}
\par\end{centering}
\begin{centering}
\subfloat[]{\begin{centering}
\includegraphics[scale=0.62]{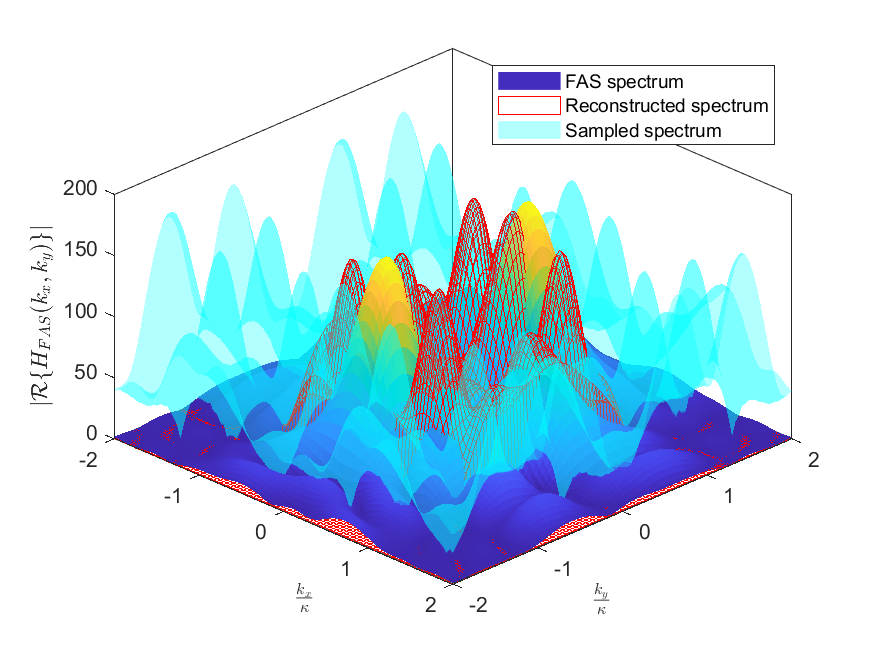} 
\par\end{centering}
}
\par\end{centering}
\centering{}\caption{The 2D power spectrum density of $\hat{h}_{{\rm FAS}}\left(x,y\right)$
using: a) $D_{0}^{\star}$ sampling distance; b) $D=\frac{\lambda}{2}$
sampling distance.}
\label{fig:2D_Recon_spectrum} \vspace{-4mm}
 
\end{figure}

Fig.~\ref{fig:2D_Recon_spatial} shows the 2D channel estimation
and reconstruction of $h_{{\rm FAS}}\left(x,y\right)$ using different
values of $D$. Similarly, we only focus on the real part as similar
behaviors can be observed in the imaginary part. As it is illustrated,
the FAS channel can be well estimated at each sampling point and efficiently
reconstructed using the suboptimal sampling distance $D_{0}^{\star}$.
This is because the effect of noise can be averaged out through the
MLE, while the suboptimal sampling distance is capable of capturing
most of the energy of the spectrum. The latter can be verified by
further analyzing the power spectrum density as shown in Fig.~\ref{fig:2D_Recon_spectrum}.
When using the sampling distance $D_{0}^{\star}$, it is observed
that the sampled and reconstructed spectra are able to capture the
spectral leakage of the mainlobe. Nevertheless, a crucial part of
spectral leakage of the mainlobe cannot be captured when using sampling
distance of $D=\frac{\lambda}{2}$, leading to inaccurate channel
reconstruction. For example, due to severe aliasing, we can see that
$D=\frac{\lambda}{2}$ sampling distance has more compact repetition
in the sampled spectrum than $D_{0}^{\star}$ sampling distance. Moreover,
the reconstructed spectrum of $D=\frac{\lambda}{2}$ sampling distance
does not fit well with the FAS spectrum.

\begin{figure}
\begin{centering}
\includegraphics[scale=0.62]{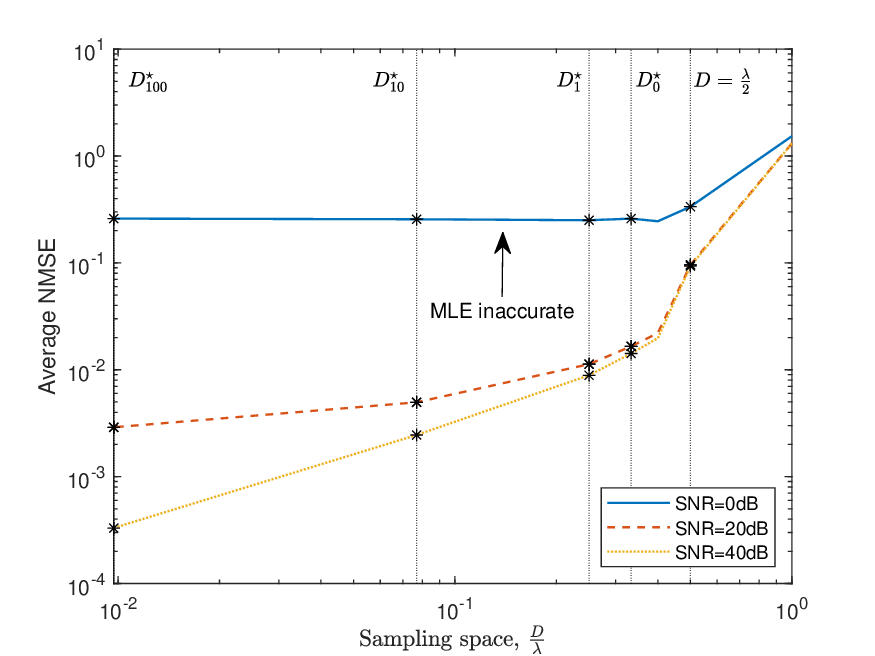} 
\par\end{centering}
\centering{}\caption{The tradeoff between the average NMSE and sampling distance, $D$.}
\label{fig:spacing} 
\end{figure}

Fig.~\ref{fig:spacing} presents the tradeoff between the average
NMSE and sampling distance, $D$. The sampling distance is closely
related to the number of estimated channels, as revealed in (\ref{eq:44})
and (\ref{eq:45}). Since $\left|H_{{\rm {FAS}}}\left(k_{x},k_{y}\right)\right|$
spans over the entire domain, the average NMSE generally decreases
as the sampling distance decreases. However, if the MLE is inaccurate,
decreasing the sampling distance may not decrease the average NMSE
as shown in the low SNR case. Moreover, it is seen that if the MLE
is accurate, then the decrease of the average NMSE is almost concave
in a log-log scale when the sampling distance is larger than $D_{0}^{\star}$
and in contrast, the decrement of the average NMSE is approximately
convex when the sampling distance is smaller than $D_{0}^{\star}$.
Although smaller sampling distance decreases the average NMSE, it
is worth noting that a smaller sampling distance comes with implementation
complexity since it could lead to frequent position switching for
liquid-based fluid antenna or strong mutual coupling for pixel-based
fluid antenna. Moreover, it would also cause significant rate degradation.
Thus, we suggest using the suboptimal sampling distance $D_{0}^{\star}$
as it is the largest sampling distance that at least covers the mainlobe
of $\left|H_{{\rm {FAS}}}\left(k_{x},k_{y}\right)\right|$.

\begin{figure}
\begin{centering}
\subfloat[]{\centering{}\includegraphics[scale=0.62]{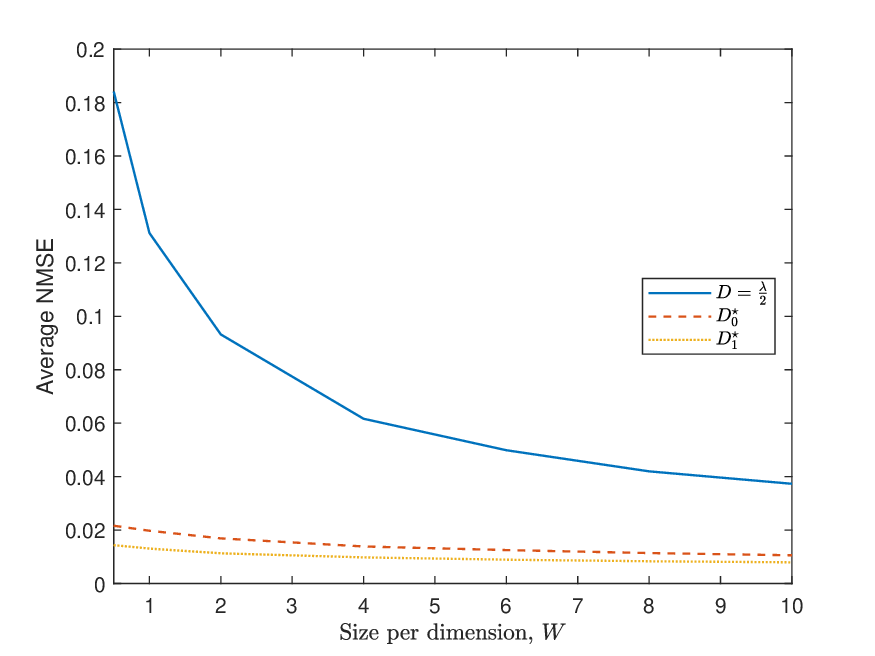} 

}
\par\end{centering}
\begin{centering}
\subfloat[]{\centering{}\includegraphics[scale=0.62]{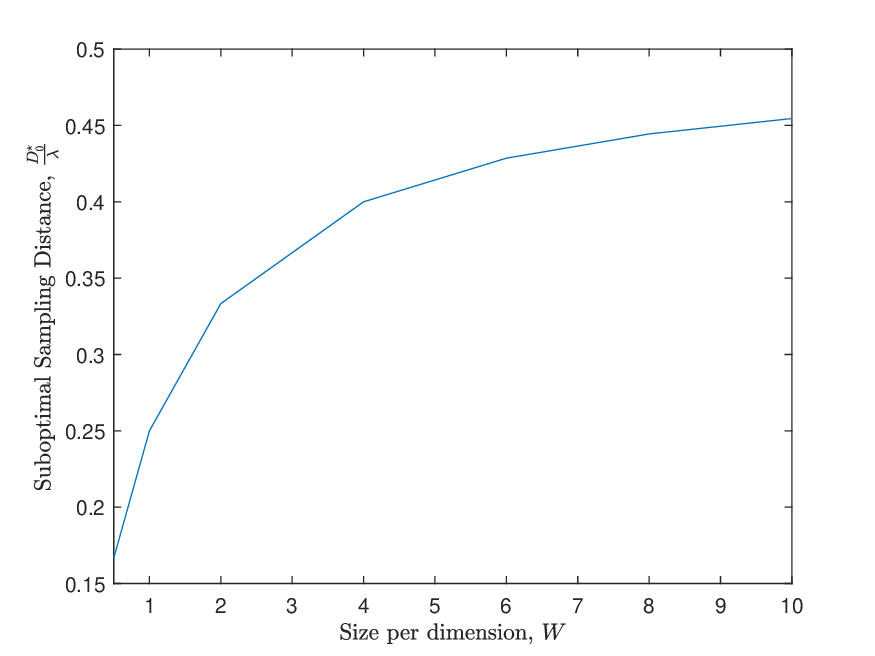} 

}
\par\end{centering}
\centering{}\caption{The sensitivity of sampling distance: a) the average NMSE versus $W$;
b) $D_{0}^{\star}$ versus $W$.}
\label{fig:length} \vspace{-4mm}
 
\end{figure}

Fig.~\ref{fig:length}(a) highlights the sensitivity of the sampling
distance over different values of $W$ in terms of the average NMSE.
As anticipated, the average NMSE of $D=\frac{\lambda}{2}$ is evidently
higher than that of $D_{0}^{\star}$ when $W$ is small. We may think
that the average NMSE of $D=\frac{\lambda}{2}$ may converge to that
of $D_{0}^{\star}$ when $W$ is large. Nevertheless, our results
show that there is still a large gap between $D_{0}^{\star}$ and
$D=\frac{\lambda}{2}$ even when $W$ is increased to $10$. The reason
is that $D_{0}^{\star}$ only converges to $\frac{\lambda}{2}$ at
a rate proportional to $\frac{1}{W}$, as seen in Fig.~\ref{fig:length}(b).
Thus, extremely large $W$ is still required to make $\frac{\lambda}{2}$
sufficient for an efficient channel reconstruction. This suggests
that a minor variation in the sampling distance can affect the average
NMSE substantially when $W$ is large. Note that when $W$ is extremely
large, the FAS receiver might be located at the near-field region.
In that case, a smaller sampling distance is always useful \cite{di2023electromagnetic}.

\begin{figure}
\begin{centering}
\subfloat[]{\begin{centering}
\includegraphics[scale=0.62]{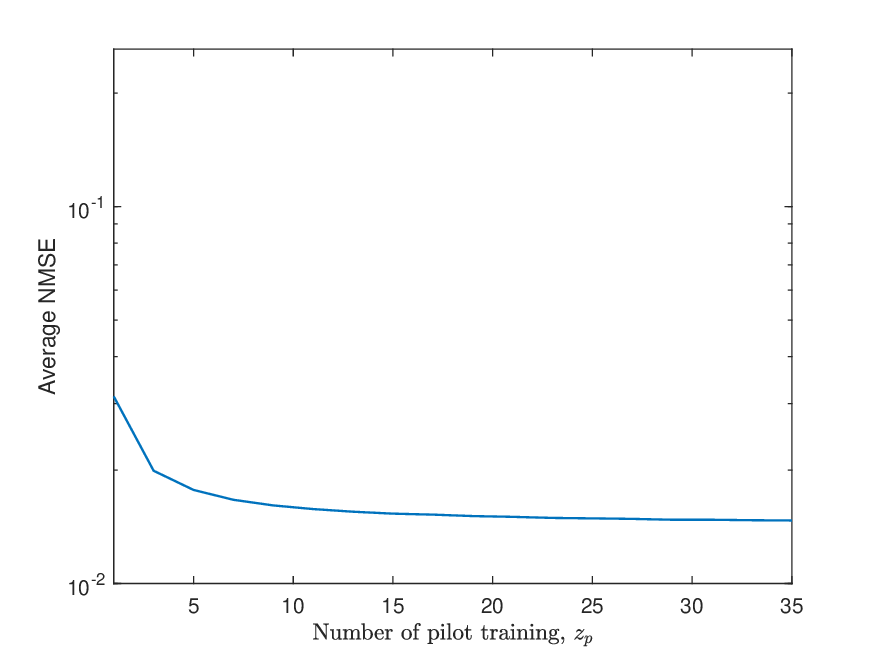} 
\par\end{centering}
}
\par\end{centering}
\begin{centering}
\subfloat[]{\begin{centering}
\includegraphics[scale=0.62]{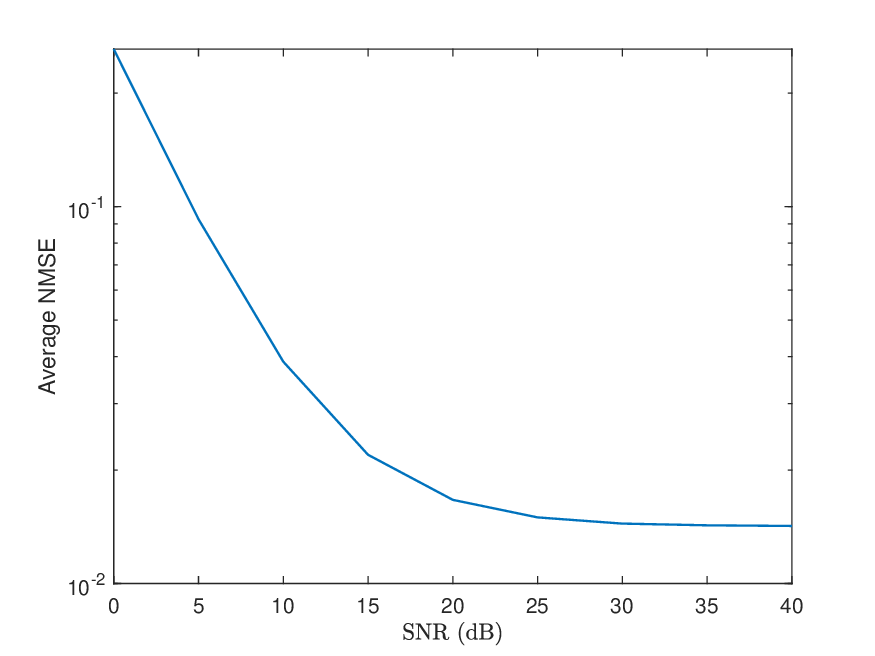} 
\par\end{centering}
}
\par\end{centering}
\centering{}\subfloat[]{\begin{centering}
\includegraphics[scale=0.62]{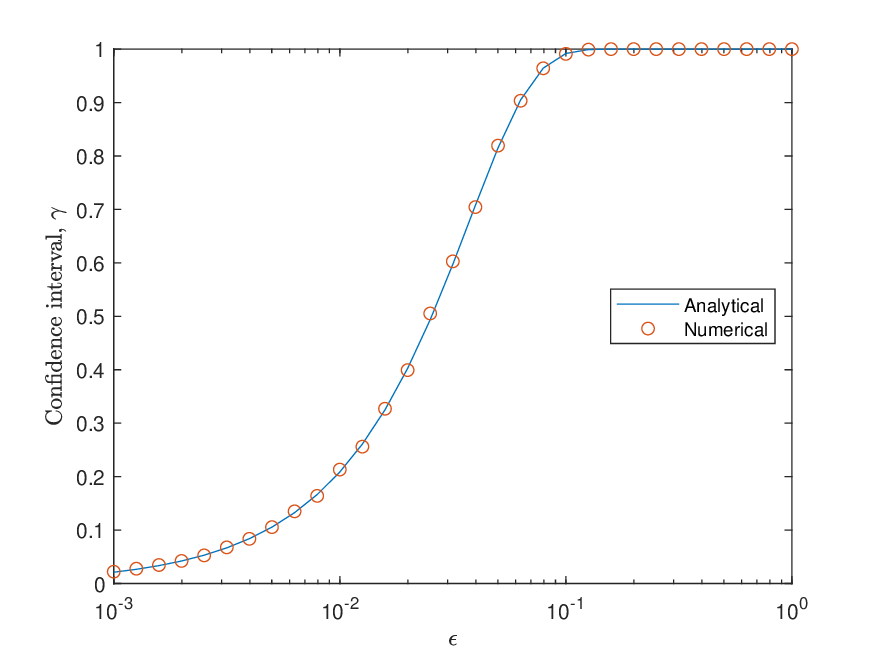} 
\par\end{centering}
}\caption{The effects of MLE based on different values of: a) $z_{p}$; b) ${\rm SNR}$;
and c) $\varepsilon$.}
\label{fig:MLE} \vspace{-4mm}
 
\end{figure}

Fig.~\ref{fig:MLE} demonstrates the effects of $z_{p}$, ${\rm SNR}$,
and $\varepsilon$. As it is seen, the average NMSE always decreases
as $z_{p}$ or SNR increases. However, the effect of $z_{p}$ is less
substantial than the SNR. This is because in practice, $z_{p}$ could
only be adjusted linearly while the ${\rm SNR}$ usually ranges in
the logarithmic scale. Furthermore, the numerical CI closely matches
the analytical CI, and thus we can guarantee that the channel estimation
error is bounded by $\varepsilon$ with a tractable CI, $\gamma$,
implying that we can determine the value of $z_{p}^{*}$ given some
fixed $\varepsilon$, ${\rm SNR}$, and $\gamma$-CI.

\begin{figure}[t]
\begin{centering}
\includegraphics[scale=0.62]{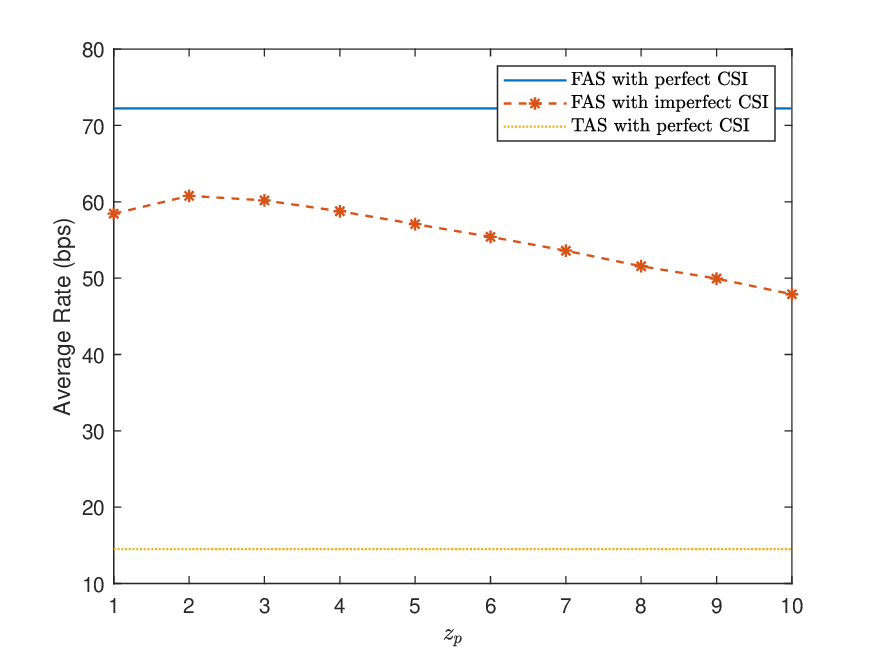} 
\par\end{centering}
\centering{}\caption{The rates of FAS and TAS versus $z_{p}$.}
\label{fig:rate} \vspace{-4mm}
 
\end{figure}

Fig.~\ref{fig:rate} evaluates the rates of FAS and TAS over different
values of $z_{p}$. To highlight the tradeoff between high data rate
and accurate channel estimation and reconstruction, we fix the SNR
to $10~{\rm dB}$. For ease of exposition, we represent the proposed
Nyquist sampling and MLE methods as FAS with imperfect CSI. As observed,
FAS with perfect CSI delivers the highest data rate, followed by FAS
with imperfect CSI, and finally TAS with perfect CSI. This result
shows that on one hand, FAS can significantly outperform TAS while
considering the channel estimation and reconstruction problem. On
the other hand, there is an optimal value of $z_{p}$ that maximizes
the rate of FAS with imperfect CSI. Nevertheless, the optimal $z_{p}$
that maximizes the achievable rate intricately depends on the system
parameters $D,W,Z,$ and ${\rm SNR}$. Note that FAS can outperform
TAS with perfect CSI because it can exploit the spatial diversity
within a given space, while TAS cannot \cite{new2024tutorial}.

\begin{figure}[t]
\begin{centering}
\includegraphics[scale=0.62]{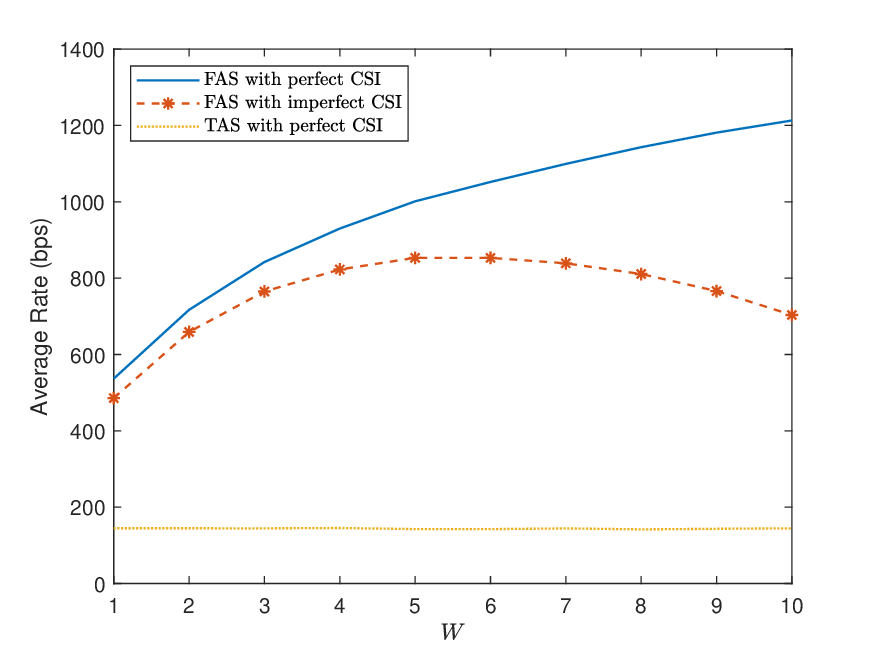} 
\par\end{centering}
\centering{}\caption{The rates of FAS and TAS versus $W$.}
\label{fig:rate-size} \vspace{-4mm}
 
\end{figure}

Fig.~\ref{fig:rate-size} compares the rate of FAS and TAS across
different values of $W$. While existing studies suggest that increasing
$W$ is preferable for improving the FAS performance, thanks to the
additional spatial diversity gain, the consideration of channel estimation
alters this conclusion. Specifically, there is an optimal $W$ that
maximizes the achievable rate of FAS. This is because a larger $W$
requires more estimated channels to efficiently reconstruct the FAS
channel, which can lead to rate degradation. Therefore, a smaller
$W$ or few estimated channels may be more promising in practice.

\begin{rem}
{\em Our studies discover that the initial discrepancies stem from
subtle differences in the research focus. In fields such as holographic
MIMO or electromagnetic information theory, the primary concern is
maximizing transmission performance. This focus often leads to the
utilization of the entire continuous antenna aperture for transmission,
enabling researchers to concentrate exclusively on the electromagnetic
field, as the channel properties become the limiting factor for transmission
efficiency. In contrast, FAS typically does not employ the entire
antenna aperture for transmission. However, obtaining the full CSI
is crucial to optimizing the performance of FAS. Since a FAS receiver
only observes the electromagnetic field across a finite space, the
received signal is not band-limited. Moreover, the channel itself
might not be inherently band-limited in practice. Therefore, oversampling
is essential for accurate channel reconstruction.} 
\end{rem}

\section{Conclusion}

\label{sec:conclude} In this paper, we investigated the problem of
channel estimation and reconstruction in FAS using the Nyquist sampling
and MLE methods. In contrast to prior studies, we incorporated an
electromagnetic-compliant channel model to enhance our understanding
in this problem. Our analysis indicated that using a half-wavelength
sampling distance, though fundamental, proved inefficient in terms
of NMSE due to the limited spatial reception of the FAS receiver.
This spatial constraint necessitated oversampling to achieve accurate
channel reconstruction, establishing a critical tradeoff between the
accuracy of the reconstructed channel and the number of estimated
channels. To strike a balance between these metrics, we derived closed-form
expressions for the suboptimal sampling distance and the total number
of estimated channels needed for a given space. This suboptimal approach
was designed to capture the majority of the energy from the mainlobe
or $d$-th sidelobe of the channel spectrum. Additionally, we employed
MLE to estimate the channel at predetermined locations, ensuring that
the channel estimation error remained within a bound of $\varepsilon$,
with a tractable CI. Our analysis further revealed the relationships
between $\varepsilon$, $z_{p}$, ${\rm SNR}$, and $\gamma$-CI.
Utilizing these techniques allowed us to effectively determine the
total number of pilot symbols required for the channel estimation
and reconstruction process in FAS. Finally, we demonstrated that FAS
with imperfect CSI can significantly outperform TAS with perfect CSI
using the proposed Nyquist sampling and MLE methods. Furthermore,
there exists an optimal $W$ that maximizes the rate of FAS with imperfect
CSI, as increasing $W$ requires more estimated channels, potentially
leading to rate degradation.

\section*{Appendix I: 1D Nyquist Sampling Method }

Let us denote the spatial spectrum of $h\left(x\right)$ as $H\left(k_{x}\right)$.
In a 1D fluid antenna surface, the receiver can only observe $h\left(x\right)$
over a finite length of $X\lambda$. Thus, it is necessary to consider
the following 1D rectangular window function 
\begin{equation}
u\left(x\right)=\begin{cases}
1, & -\frac{X}{2}\lambda\leq x\leq\frac{X}{2}\lambda,\\
0, & {\rm {otherwise}.}
\end{cases}
\end{equation}

\noindent The channel observed at the FAS receiver is given by 
\begin{equation}
h_{{\rm FAS}}\left(x\right)=h\left(x\right)u\left(x\right).
\end{equation}
In the frequency domain, the spectrum of $h_{{\rm FAS}}\left(x\right)$
can be expressed as 
\begin{equation}
H_{{\rm FAS}}\left(k_{x}\right)=H\left(k_{x}\right)\circledast U\left(k_{x}\right),
\end{equation}
where 
\begin{equation}
U\left(k_{x}\right)=\frac{\sin\left(\frac{k_{x}}{2}X\lambda\right)}{\frac{k_{x}}{2}}.
\end{equation}

\noindent Apparently, $H_{{\rm FAS}}\left(k_{x}\right)$ is no longer
band-limited. Using similar steps as the 2D Nyquist sampling method,
we can obtain the $d$-th efficient sampling distance and the minimum
number of estimated channels as 
\begin{equation}
D_{d}^{\star}=\frac{\lambda}{2+\frac{2}{X}\left(d+1\right)}.
\end{equation}
and 
\begin{equation}
N_{d}^{\star}=\left\lfloor \frac{X\lambda}{D_{d}^{\star}}\right\rfloor ,
\end{equation}
respectively.

The sampled signal in the 1D case can be expressed as 
\begin{equation}
h_{{\rm FAS}}^{s}\left(x\right)=\sum_{n=0}^{N_{d}^{\star}}\hat{h}_{{\rm FAS}}\left(nD_{d}^{\star}\right)\delta\left(x-nD_{d}^{\star}\right).
\end{equation}
In the frequency domain, we hence have 
\begin{align}
H_{{\rm FAS}}^{s}\left(k\right)= & \sum_{n=0}^{N_{d}^{\star}}\hat{h}_{{\rm FAS}}\left(nD_{d}^{\star}\right)e^{-jknD_{d}^{\star}}\\
= & \frac{1}{D_{d}^{\star}}\sum_{n=0}^{N_{d}^{\star}}H_{{\rm FAS}}\left(k-nk_{s}^{\star}\right),
\end{align}
where $k_{s}^{\star}=\frac{2\pi}{D_{d}^{\star}}$ denotes the sampling
frequency. Furthermore, $\hat{h}_{{\rm FAS}}\left(x\right)$ can be
reconstructed using a low-pass filter as 
\begin{equation}
\hat{H}_{{\rm FAS}}\left(k\right)=H_{{\rm FAS}}^{s}\left(k\right)f\left(k\right),
\end{equation}
where 
\begin{equation}
f\left(k\right)=\begin{cases}
1, & -\frac{2\pi}{D_{d}^{\star}}\leq k\leq\frac{2\pi}{D_{d}^{\star}},\\
0, & {\rm {otherwise}.}
\end{cases}
\end{equation}
Finally, we can obtain $\hat{h}_{{\rm FAS}}\left(x\right)$ for $-\frac{X}{2}\lambda\leq x\leq\frac{X}{2}\lambda$
using the inverse Fourier transform as 
\begin{equation}
\hat{h}_{{\rm FAS}}\left(x\right)=\frac{1}{2\pi}\int\hat{H}_{{\rm FAS}}\left(k\right)e^{-jk_{x}x}dx.
\end{equation}
The NMSE between $\hat{h}_{{\rm FAS}}\left(x\right)$ and $h_{{\rm FAS}}\left(x\right)$
is simplified as 
\begin{equation}
{\rm {NMSE}}=\frac{\int_{-\frac{X}{2}\lambda}^{\frac{X}{2}\lambda}\left(\hat{h}_{{\rm FAS}}\left(x\right)-h_{{\rm FAS}}\left(x\right)\right)^{2}dx}{\int_{-\frac{X}{2}\lambda}^{\frac{X}{2}\lambda}\left(h_{{\rm FAS}}\left(x\right)\right)^{2}dx}.
\end{equation}


\begin{thebibliography}{10}
\bibitem{dash2023selection} S.~Dash, C.~Psomas, and I.~Krikidis,
``Selection of metallic liquid in {sub-6 GHz} antenna design for
{6G} networks,'' {\em Scientific Reports}, vol.~13, no.~1,
p.~20551, Nov. 2023.

\bibitem{9899974} Y.~Shen, K.-F. Tong, and K.-K. Wong, ``Radiation
pattern diversified single-fluid-channel surface-wave antenna for
mobile communications,'' in {\em Proc. IEEE-APS Topical Conf. Antennas
\& Propag. Wireless Commun.}, pp.~49--51, 5-9 Sept. 2022, Cape
Town, South Africa.

\bibitem{1367557} B.~Cetiner {\em et al.}, ``Multifunctional
reconfigurable {MEMS} integrated antennas for adaptive {MIMO}
systems,'' {\em IEEE Commun. Mag.}, vol.~42, no.~12, pp.~62--70,
Dec. 2004.

\bibitem{9785489} L.~Jing, M.~Li, and R.~Murch, ``Compact pattern
reconfigurable pixel antenna with diagonal pixel connections,'' {\em
IEEE Trans. Antennas \& Propag.}, vol.~70, no.~10, pp.~8951--8961,
Oct. 2022.

\bibitem{8060521} S.~Basbug, ``Design and synthesis of antenna
array with movable elements along semicircular paths,'' {\em IEEE
Antennas \& Wireless Propag. Lett.}, vol.~16, pp.~3059--3062,
Oct. 2017.

\bibitem{Wong-ell2020} K.-K. Wong, K.-F. Tong, Y.~Zhang, and Z.~Zheng,
``Fluid antenna system for {6G}: When {Bruce Lee} inspires wireless
communications,'' {\em Elect. Lett.}, vol.~56, no.~24, pp.~1288--1290,
Nov. 2020.

\bibitem{wong2022bruce} K.-K. Wong, K.-F. Tong, Y.~Shen, Y.~Chen,
and Y.~Zhang, ``{Bruce Lee}-inspired fluid antenna system: Six
research topics and the potentials for {6G},'' {\em Frontiers
Commun. Netw.}, vol. 3, no. 853416, Mar. 2022.

\bibitem{9770295} A.~Shojaeifard {\em et al.}, ``{MIMO} evolution
beyond {5G} through reconfigurable intelligent surfaces and fluid
antenna systems,'' {\em Proc. IEEE}, vol.~110, no.~9, pp.~1244--1265,
Sept. 2022.

\bibitem{10480333} J.~Zheng {\em et al.}, ``Flexible-position
{MIMO} for wireless communications: Fundamentals, challenges, and
future directions,'' {\em IEEE Wireless Commun.}, early access,
\url{doi:10.1109/MWC.011.2300428}, Mar. 2024.

\bibitem{Shen-tap_submit2024} Y. Shen {\em et al.}, ``Design
and implementation of mmWave surface wave enabled fluid antennas and
experimental results for fluid antenna multiple access,'' {\em
arXiv preprint}, \url{arXiv:2405.09663}, May 2024. 

\bibitem{Zhang-pFAS2024} J.~Zhang {\em et al.}, ``A pixel-based
reconfigurable antenna design for fluid antenna systems,'' \emph{arXiv
preprint}, \url{arXiv:2406.05499}, Jun. 2024.

\bibitem{9264694} K.-K. Wong, A.~Shojaeifard, K.-F. Tong, and Y.~Zhang,
``Fluid antenna systems,'' {\em IEEE Trans. Wireless Commun.},
vol.~20, no.~3, pp.~1950--1962, Mar. 2021.

\bibitem{10103838} M.~Khammassi, A.~Kammoun, and M.-S. Alouini,
``A new analytical approximation of the fluid antenna system channel,''
{\em IEEE Trans. Wireless Commun.}, vol. 22, no. 12, pp. 8843--8858,
Dec. 2023.

\bibitem{ramirez2024new} P.~Ram�rez-Espinosa, D.~Morales-Jimenez,
and K.-K. Wong, ``A new spatial block-correlation model for fluid
antenna systems,'' {\em IEEE Trans. Wireless Communications},
early access, \url{doi:10.1109/TWC.2024.3434509}, Aug. 2024.

\bibitem{10130117} W.~K. New, K.-K. Wong, H.~Xu, K.-F. Tong, and
C.-B. Chae, ``Fluid antenna system: New insights on outage probability
and diversity gain,'' {\em IEEE Trans. Wireless Commun.}, vol.
23, no. 1, pp. 128--140, Jan. 2024.

\bibitem{alvim2023performance} P.~D. Alvim {\em et al.}, ``On
the performance of fluid antennas systems under $\alpha$-$\mu$ fading
channels,'' {\em IEEE Wireless Commun. Lett.}, vol. 13, no. 1,
pp. 108--112, Jan. 2024.

\bibitem{lopez2023novel} J.~D. Vega-S{a}nchez, A.~E. L{o}pez-Ram{i}rez,
L.~Urquiza-Aguiar, and D.~P.~M. Osorio, ``{Novel expressions
for the outage probability and diversity gains in fluid antenna system},''
\emph{IEEE Wirel. Commun. Lett.}, vol. 13, no. 2, pp. 372--376, Feb.
2024

\bibitem{10303274} W. K. New, K. K. Wong, H. Xu, K. F. Tong and C.-B.
Chae, ``An information-theoretic characterization of MIMO-FAS: Optimization,
diversity-multiplexing tradeoff and $q$-outage capacity,'' {\em
IEEE Trans. Wireless Commun.}, vol. 23, no. 6, pp. 5541--5556, Jun.
2024.

\bibitem{10354059} Y.~Chen, S.~Li, Y.~Hou, and X.~Tao, ``Energy-efficiency
optimization for slow fluid antenna multiple access using mean-field
game,'' {\em IEEE Wireless Commun. Lett.}, vol. 13, no. 4, pp.
915--918, Apr. 2024.

\bibitem{10354003} L.~Zhu, W.~Ma, B.~Ning, and R.~Zhang, ``Movable-antenna
enhanced multiuser communication via antenna position optimization,''
{\em IEEE Trans. Wireless Commun.}, vol. 23, no. 7, pp. 7214--7229,
Jul. 2024.

\bibitem{xu2023capacity} H.~Xu {\em et al.}, ``Capacity maximization
for {FAS}-assisted multiple access channels,'' {\em arXiv preprint},
\url{arXiv:2311.11037}, 2023.

\bibitem{ISAC_FAS} C.~Wang {\em et al.}, ``Fluid antenna system
liberating multiuser {MIMO} for {ISAC} via deep reinforcement
learning,'' {\em IEEE Trans. Wireless Commun.}, early access,
\url{doi:10.1109/TWC.2024.3376800}, Mar. 2024.

\bibitem{ghadi2024performance} F.~R. Ghadi {\em et al.}, ``On
performance of {RIS}-aided fluid antenna systems,'' {\em IEEE
Wireless Commun. Lett.}, vol. 13, no. 8, pp. 2175--2179, Aug. 2024.

\bibitem{10092780} B.~Tang {\em et al.}, ``Fluid antenna enabling
secret communications,'' {\em IEEE Commun. Lett.}, vol.~27, no.~6,
pp.~1491--1495, Jun. 2023. 

\bibitem{Osorio-tvt2024} J. D. Vega-S{a}nchez, L. Urquiza-Aguiar,
H. R. C. Mora, N. V. O. Garz{o}n and D. P. M. Osorio, ``Fluid antenna
system: Secrecy outage probability analysis,'' {\em IEEE Trans.
Veh. Technol.}, vol. 73, no. 8, pp. 11458--11469, Aug. 2024.

\bibitem{10167904} L.~Tlebaldiyeva, S.~Arzykulov, T.~A. Tsiftsis,
and G.~Nauryzbayev, ``Full-duplex cooperative {NOMA}-based {mmWave}
networks with fluid antenna system ({FAS}) receivers,'' in {\em
Proc. Inter. Balkan Conf. Commun. Netw.}, 5-8 Jun. 2023, Istanbul,
Turkey. 

\bibitem{10318134} W.~K. New {\em et al.}, ``Fluid antenna system
enhancing orthogonal and non-orthogonal multiple access,'' {\em
IEEE Commun. Lett.}, vol. 28, no. 1, pp. 218--222, Jan. 2024.

\bibitem{10184308} C.~Skouroumounis and I.~Krikidis, ``Fluid antenna-aided
full duplex communications: A macroscopic point-of-view,'' {\em
IEEE J. Select. Areas Commun.}, vol.~41, no.~9, pp.~2879--2892,
Sept. 2023.

\bibitem{zhu2023index} J.~Zhu {\em et al.}, ``Index modulation
for fluid antenna-assisted {MIMO} communications: System design
and performance analysis,'' {\em IEEE Trans. Wireless Commun.},
vol. 23, no. 8, pp. 9701--9713, Aug. 2024. 

\bibitem{Xu-2024} Y. Chen, and T. Xu, ``Fluid antenna index modulation
communications,'' {\em IEEE Wireless Commun. Lett.}, vol. 13,
no. 4, pp. 1203--1207, Apr. 2024. 

\bibitem{Yang-2024} H. Yang {\em et al.}, ``Position index modulation
for fluid antenna system,'' {\em IEEE Trans. Wireless Commun.},
early access, \url{doi:10.1109/TWC.2024.3446658}, Aug. 2024.

\bibitem{zhu2024historical} L.~Zhu and K. K. Wong, ``Historical
review of fluid antenna and movable antenna,'' {\em arXiv preprint},
\url{arXiv:2401.02362v2}, 2024.

\bibitem{8794743} M.~Wang, F.~Gao, S.~Jin, and H.~Lin, ``An
overview of enhanced massive {MIMO} with array signal processing
techniques,'' {\em IEEE J. Select. Topics Sig. Process.}, vol.~13,
no.~5, pp.~886--901, Sept. 2019.

\bibitem{8272484} D.~Neumann, T.~Wiese, and W.~Utschick, ``Learning
the {MMSE} channel estimator,'' {\em IEEE Trans. Sig. Process.},
vol.~66, no.~11, pp.~2905--2917, Jun. 2018.

\bibitem{9931521} L.~Cheng, F.~Yin, S.~Theodoridis, S.~Chatzis,
and T.-H. Chang, ``Rethinking {Bayesian} learning for data analysis:
The art of prior and inference in sparsity-aware modeling,'' {\em
IEEE Sig. Process. Mag.}, vol.~39, no.~6, pp.~18--52, Nov. 2022.

\bibitem{5954192} M.~F. Duarte and Y.~C. Eldar, ``Structured compressed
sensing: From theory to applications,'' {\em IEEE Trans. Sig. Process.},
vol.~59, no.~9, pp.~4053--4085, Sept. 2011.

\bibitem{9732214} G.~Zhou, C.~Pan, H.~Ren, P.~Popovski, and A.~L.
Swindlehurst, ``Channel estimation for {RIS}-aided multiuser millimeter-wave
systems,'' {\em IEEE Trans. Sig. Process.}, vol.~70, pp.~1478--1492,
Mar. 2022.

\bibitem{9180053} G.~Zhou, C.~Pan, H.~Ren, K.~Wang, and A.~Nallanathan,
``A framework of robust transmission design for {IRS}-aided {MISO}
communications with imperfect cascaded channels,'' {\em IEEE Trans.
Sig. Process.}, vol.~68, pp.~5092--5106, Aug. 2020.

\bibitem{10262375} Z.~Peng {\em et al.}, ``Two-stage channel
estimation for {RIS}-aided multiuser {mmWave} systems with reduced
error propagation and pilot overhead,'' {\em IEEE Trans. Sig. Process.},
vol.~71, pp.~3607--3622, Sept. 2023.

\bibitem{10495003} H.~Zhang {\em et al.}, ``Learning-induced
channel extrapolation for fluid antenna systems using asymmetric graph
masked autoencoder,'' {\em IEEE Wireless Commun. Lett.}, vol.
13, no. 6, pp. 1665--1669, Jun. 2024. 

\bibitem{10018377} N.~Waqar, K.-K. Wong, K.-F. Tong, A.~Sharples,
and Y.~Zhang, ``Deep learning enabled slow fluid antenna multiple
access,'' {\em IEEE Commun. Lett.}, vol.~27, no.~3, pp.~861--865,
Mar. 2023. 

\bibitem{Eskandari-2024} M. Eskandari, A. Burr, K. Cumanan, and K.
K. Wong, ``cGAN-based slow fluid antenna multiple access,'' accepted
in {\em IEEE Wireless Commun. Lett.}, 2024.

\bibitem{10375559} H.~Xu {\em et al.}, ``Channel estimation
for FAS-assisted multiuser mmWave systems,'' {\em IEEE Commun.
Lett.}, vol. 23, no. 3, pp. 632--636, Mar. 2024.

\bibitem{zhang2024successive} Z.~Zhang, J.~Zhu, L.~Dai, and R.~W.
Heath~Jr, ``Successive {Bayesian} reconstructor for channel estimation
in fluid antenna systems,'' in Proc. {\em IEEE Wireless Commun.
Netw. Conf.}, 21-24 Apr. 2024, Dubai, United Arab Emirates.

\bibitem{zhang2023successive} Z.~Zhang, J.~Zhu, L.~Dai, and R.~W.
Heath~Jr, ``Successive {Bayesian} reconstructor for channel estimation
in fluid antenna systems,'' {\em arXiv preprint}, \url{arXiv:2312.06551},
2023.

\bibitem{10236898} W.~Ma, L.~Zhu, and R.~Zhang, ``Compressed
sensing based channel estimation for movable antenna communications,''
{\em IEEE Commun. Lett.}, vol.~27, no.~10, pp.~2747--2751,
Oct. 2023.

\bibitem{zhu2022electromagnetic} J.~Zhu, Z.~Wan, L.~Dai, M.~Debbah,
and H.~V. Poor, ``Electromagnetic information theory: Fundamentals,
modeling, applications, and open problems,'' {\em IEEE Wireless
Commun.}, vol. 31, no. 3, pp. 156--162, Mar. 2024.

\bibitem{bjornson2024towards} E.~Bj{o}rnson {\em et al.}, ``Towards
{6G MIMO}: Massive spatial multiplexing, dense arrays, and interplay
between electromagnetics and processing,'' {\em arXiv preprint},
\url{arXiv:2401.02844}, 2024.

\bibitem{9798854} A.~Pizzo, A.~d.~J. Torres, L.~Sanguinetti,
and T.~L. Marzetta, ``Nyquist sampling and degrees of freedom of
electromagnetic fields,'' {\em IEEE Trans. Sig. Process.}, vol.~70,
pp.~3935--3947, Jun. 2022.

\bibitem{di2023electromagnetic} M.~Di~Renzo and M.~D. Migliore,
``Electromagnetic signal and information theory--on electromagnetically
consistent communication models for the transmission and processing
of information,'' {\em arXiv preprint}, \url{arXiv:2311.06661},
2023.

\bibitem{8437634} T.~L. Marzetta, ``Spatially-stationary propagating
random field model for massive {MIMO} small-scale fading,'' in
Proc. {\em IEEE Int. Symp. Inform. Theory}, pp.~391--395, 17-22
Jun. 2018, Vail, Colorado, USA.

\bibitem{9110848} A.~Pizzo, T.~L. Marzetta, and L.~Sanguinetti,
``Spatially-stationary model for holographic {MIMO} small-scale
fading,'' {\em IEEE J. Select. Areas Commun.}, vol.~38, no.~9,
pp.~1964--1979, Sept. 2020.

\bibitem{hildebrand1962advanced} F.~B. Hildebrand, {\em Advanced
calculus for applications}, Pearson, 2nd Edition, 1962.

\bibitem{balakrishnan2003all} V.~Balakrishnan, ``All about the
{Dirac} delta function (?),'' {\em Resonance}, vol.~8, no.~8,
pp.~48--58, 2003.

\bibitem{1697831} C.~Shannon, ``Communication in the presence of
noise,'' {\em Proc. IRE}, vol.~37, no.~1, pp.~10--21, 1949.

\bibitem{landau1985overview} H.~Landau, ``An overview of time and
frequency limiting,'' {\em Fourier Tech. \& Appl.}, pp.~201--220,
1985.

\bibitem{lathi2005linear} B.~P. Lathi and R.~A. Green, {\em Linear
systems and signals}, vol.~2, Oxford University Press New York,
2005.

\bibitem{841172} M.~Medard, ``The effect upon channel capacity
in wireless communications of perfect and imperfect knowledge of the
channel,'' {\em IEEE Trans. Inform. Theory}, vol.~46, no.~3,
pp.~933--946, May 2000.

\bibitem{720551} E.~Biglieri, J.~Proakis, and S.~Shamai, ``Fading
channels: Information-theoretic and communications aspects,'' {\em
IEEE Trans. Inform. Theory}, vol.~44, no.~6, pp.~2619--2692,
Oct. 1998.

\bibitem{new2024tutorial} W. K. New {\em et al.}, ``A tutorial
on fluid antenna system for 6G networks: Encompassing communication
theory, optimization methods and hardware designs,'' {\em arXiv
preprint}, \url{arXiv:2407.03449}, Jul. 2024. 

\end{thebibliography}
\end{document}